\documentclass[amsmath,amssymb,a4paper,11pt]{article}
\usepackage[top=70pt,bottom=100pt,left=80pt,right=80pt]{geometry}

\DeclareFontFamily{OT1}{pzc}{}
\DeclareFontShape{OT1}{pzc}{m}{it}{<-> s * [1.10] pzcmi7t}{}
\DeclareMathAlphabet{\mathpzc}{OT1}{pzc}{m}{it}

\usepackage{authblk}
\usepackage{abstract}
\usepackage[footnotesize]{caption}
\usepackage{subcaption}

\usepackage{cite}
\usepackage{amsmath}
\numberwithin{equation}{section}

\usepackage{amssymb}
\usepackage{braket}
\usepackage{graphicx} 
\usepackage{mathrsfs}
\usepackage{color}
\usepackage{multirow}
\usepackage{multicol}
\usepackage{blindtext}
\usepackage{upgreek}
\usepackage{float}
\usepackage{pdfpages}
\usepackage{slashed}
\usepackage{caption}
\usepackage{subcaption}
\usepackage[export]{adjustbox}
\usepackage{appendix}

\usepackage{jheppub}
\usepackage{cleveref}

\def\gsim{\mathrel{\rlap{\lower4pt\hbox{\hskip1pt$\sim$}}
    \raise1pt\hbox{$>$}}}

\title{Heavy Neutrino-Antineutrino Oscillations \\[1mm] in Quantum Field Theory}\author[a]{Stefan~Antusch~and}
\author[a]{Johannes~Rosskopp}

\affiliation[a]{ Department of Physics, University of Basel,\\ 
 Klingelbergstr.\ 82, CH-4056 Basel, Switzerland}

\emailAdd{stefan.antusch@unibas.ch}
\emailAdd{j.rosskopp@unibas.ch}

\abstract{It has been proposed that the coherent propagation of long-lived heavy neutrino mass eigenstates can lead to an oscillating rate of lepton number conserving (LNC) and violating (LNV) events, as a function of the distance between the production and displaced decay vertices. We discuss this phenomenon, which we refer to as heavy neutrino-antineutrino oscillations, in the framework of quantum field theory (QFT), using the formalism of external wave packets. General formulae for the oscillation probabilities and the number of expected events are derived and the coherence and localisation conditions that have to be satisfied in order for neutrino-antineutrino oscillations to be observable are discussed. The formulae are then applied to a low scale seesaw scenario, which features two nearly mass degenerate heavy neutrinos that can be sufficiently long lived to produce a displaced vertex when their masses are below the $W$ boson mass. The leading and next-to-leading order oscillation formulae for this scenario are derived. For an example parameter point used in previous studies, the kinematics of the considered LNC/LNV processes are simulated, to check that the coherence and localisation conditions are satisfied. Our results show that the phenomenon of heavy neutrino-antineutrino oscillations can indeed occur in low scale seesaw scenarios and that the previously used leading order formulae, derived with a plane wave approach, provide a good approximation for the considered example parameter point.}
\arxivnumber{2012.05763}

\graphicspath{{./assets/}} 
\date{}

\begin{document}
\maketitle

\section{Introduction}

After neutrino oscillations have been proposed by Pontecorvo in the late 50's \cite{Pontecorvo:1957cp}, they have lead to great insight into the (light) neutrino parameters. First introduced as neutrino-antineutrino oscillations, Maki, Nakagawa and Sakata considered oscillations into different flavours in 1962 \cite{Maki:1962mu}. Pontecorvo proposed in 1967 the possibility of solar neutrino oscillation, after which the solar neutrino problem was discovered \cite{Davis:1968cp}. In 1969, Pontecorvo and Gribov proposed neutrino flavour oscillations as a possible solution to this problem \cite{Gribov:1968kq}. Since then, precise measurements of the light neutrino oscillations resulted in experimental values for the two mass squared differences of the light neutrinos as well as of the three mixing angles of the (PMNS) lepton mixing matrix and have provided a first indication of the range of the Dirac CP phase.
\\\\
Despite this great success regarding the light neutrino parameters, the origin of the light neutrino masses, which requires an extension of the present Standard Model (SM) of elementary particles, is still unknown. One way to generate them consists in introducing right-chiral neutrinos as SM singlets. They can have Majorana mass terms as well as Yukawa couplings to the left-chiral SM neutrinos and the Higgs doublet. After electroweak symmetry breaking, the particle spectrum contains the three light neutrinos plus additional heavy neutrino mass eigenstates. If such heavy neutrinos are within reach of collider experiments, and if they are sufficiently long-lived, oscillations among the heavy neutrino and antineutrino interaction eigenstates could lead to great insight into the heavy neutrino parameters and thus into the neutrino mass generation mechanism.\footnote{\label{fn:definition_heavy_neutrino_interaction_eigentate}We define a neutrino (antineutrino) as the neutral lepton that is produced together with a charged antilepton (lepton) and a $W$ boson. The heavy neutrino (antineutrino) interaction eigenstates are defined as the projection of the neutrino (antineutrino) state onto the subspace of heavy mass eigenstates.}
\\\\
In parallel to the applications of neutrino oscillations also the framework in which they are described has evolved. The first approach to neutrino oscillations has been a quantum mechanical description in which the neutrinos are treated as plane waves. For neutral K meson oscillations, a quantum field model including wave packets was proposed in 1963 in Ref.~\cite{Sachs:1963}. In 1981, the authors of \cite{Kayser:1981ye} pointed out conceptual problems of the plane wave approach for light neutrino oscillations and suggested a wave packet treatment as solution. To resolve the remaining difficulties, \cite{Giunti:1993se} introduced a quantum field theoretical model similar to \cite{Kayser:1981ye}, in which the propagating particle (assumed stable) is treated as an internal line in a Feynman diagram and the external particles are described by wave packets. A review of the existing quantum field theoretical approaches is given in \cite{Beuthe:2001rc}, where also a framework employing external wave packets is discussed that can be used to describe the oscillations of unstable particles.
\\\\
To describe heavy neutrino-antineutrino oscillations, a density matrix formalism (based on \cite{Cohen:2008qb}) has been used in \cite{Cvetic:2015ura} for heavy neutrinos produced from meson decays. Using this formalism, formulae for the oscillation of the LNC and LNV decay rates have been calculated. The authors of \cite{Anamiati:2016uxp} used a formalism for meson oscillations and plane wave arguments to derive formulae for heavy neutrino-antineutrino oscillations. These formulae are then applied to heavy neutrinos produced from $W_R$ decays in a left-right symmetric extension of the SM, for heavy neutrino parameters testable at the LHC. There it was shown that without resolving the heavy neutrino-antineutrino oscillations, the integrated effect can induce a non-trivial ratio between LNV and LNC processes at colliders (for other early papers calculating non-trivial LNV/LNC ratios, without specifying oscillation formulae, see e.g.\ \cite{Gluza:2015goa,Dev:2015pga}). The parameter region in which such non-trivial ratios are expected have been discussed e.g.\ in \cite{Antusch:2017ebe,Drewes:2019byd}. In \cite{Antusch:2017ebe} it has been shown, using the oscillation formulae from plane wave arguments (cf.~\cite{Anamiati:2016uxp}), that the signature of heavy neutrino-antineutrino oscillations can be resolved at collider experiments, considering minimal low scale type I seesaw extensions of the SM.
\\\\
It has been shown in \cite{Antusch:2017ebe} that in the case of the low scale minimal linear seesaw model and inverse light neutrino mass hierarchy, the light neutrino mass splittings predict the heavy neutrino mass splitting, resulting in an oscillation length of order ${\cal O}(10\:\mbox{cm})$ that could be resolved e.g.\ at the (high-luminosity phase of) LHCb. Furthermore, it has been pointed out that in a realistic experimental setting, where the momenta of the heavy neutrinos are given by a distribution, reconstructing these momenta and considering the oscillations as a function of the heavy (anti)neutrino proper time is required to resolve the oscillation patterns. The method has been demonstrated for an example parameter point, consistent with the present searches and non-collider constraints \cite{Antusch:2017ebe}. 
\\\\
Currently there is an increasing interest in exploring heavy neutrino-antineutrino oscillations. For example, oscillations for rare $W$ boson decays at LHC are further studied in \cite{Cvetic:2018elt,Cvetic:2019rms}, and from tau decays in \cite{Zamora-Saa:2016ito,Zamora-Saa:2019naq} (using the formulae of \cite{Cvetic:2015ura}). The authors of \cite{Das:2017hmg} use plane wave arguments (together with a discussion of coherence conditions) to discuss the oscillations for a rather general right-handed neutrino mass matrix in a left-right symmetric extension of the SM. Discussions of how the insight into the low scale seesaw parameters from heavy neutrino-antineutrino oscillations can help to test whether the baryon asymmetry of the universe can be produced by the leptogenesis mechanism, are given e.g.\ in \cite{Antusch:2017pkq,Dev:2019rxh}. So far, no QFT treatment using external wave packets has been performed for heavy neutrino-antineutrino oscillations yet.
\\\\
The goal of this paper is to put the discussion of heavy neutrino-antineutrino oscillations on a more solid theoretical ground by treating them in the framework of QFT. We use the formalism of external wave packets (cf.\ \cite{Beuthe:2001rc}) to obtain a more fundamental derivation of the formulae for heavy neutrino-antineutrino oscillations, as well as a more fundamental discussion of the coherence and localisation conditions that have to be satisfied such that oscillations are observable. In addition, we apply our formulae to a specific realisation of the 
symmetry protected seesaw scenario (SPSS) \cite{Antusch:2015mia,Antusch:2016ejd} and expand them in small lepton number violating parameters. The paper is organised as follows: In section 2 the formalism of heavy neutrino-antineutrino oscillations is described, the oscillation formulae as well as formulae for the expected number of events are derived and the coherence and localisation conditions are discussed. Section 3 contains the applications to the specific SPSS model and section 4 the approximations of the oscillation formulae and the discussion of the LO and NLO effects. In section 5 we conclude.  

\section{QFT Formalism for Heavy Neutrino-Antineutrino Oscillations}
\label{sec:qft_formalism}
\begin{figure}[h]
\makebox[\textwidth][c]{
\begin{minipage}{1\textwidth}
\begin{subfigure}[c]{0.5\textwidth}
\begin{minipage}[c]{0.9\textwidth}
\begin{center}
\includegraphics{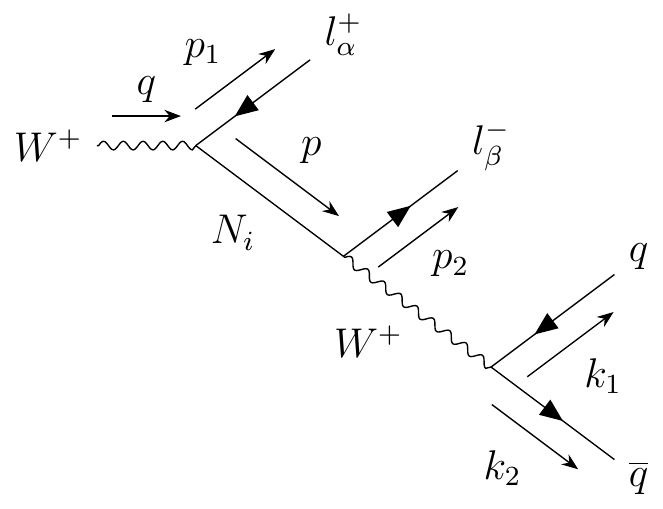}
\end{center}
\subcaption{Feynman diagram for the LNC process}
\label{feyn:LNC}
\end{minipage}
\end{subfigure}
\begin{subfigure}[c]{0.5\textwidth}
\hspace{4mm}
\begin{minipage}[c]{0.9\textwidth}
\begin{center}
 \includegraphics{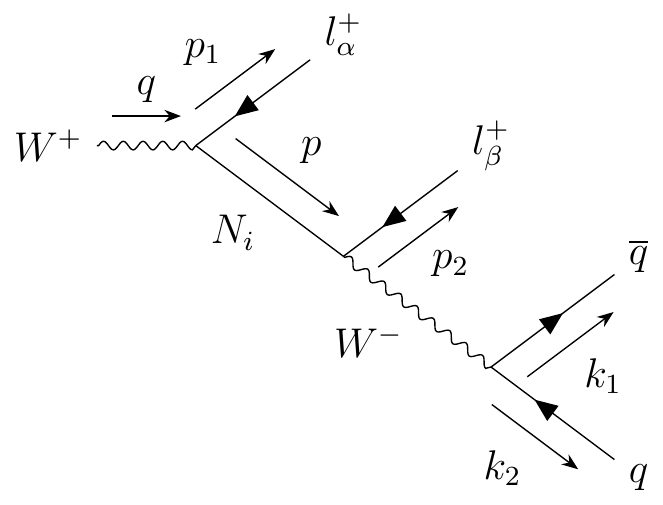}
\end{center}
\subcaption{Feynman diagram for the LNV process}
\label{feyn:LNV}
\end{minipage}
\end{subfigure}
\end{minipage}}
\caption{Feynman diagrams describing the LNC and LNV processes.}
\end{figure}

As mentioned above, we define neutrinos (antineutrinos) as those particles produced from the decay of a $W$ boson, together with a charged antilepton (lepton), respectively. As a specific example we consider in the following the dilepton-dijet signature at $pp$ colliders, which can be LNC as well as LNV. The relevant Feynman diagrams, in which the $W$ boson is produced from $pp$ collisions, are shown in \subref{feyn:LNC} and \subref{feyn:LNV}. With a $W$ boson decaying into an antilepton, the produced neutrino state is a superposition of neutrino mass eigenstates, where we only consider the heavy mass eigenstates. We define this projection of the (anti)neutrino interaction eigenstate onto the heavy neutrino mass eigenstates as ``heavy neutrino interaction eigenstate''. The neutrinos then propagate over a macroscopic distance after which they decay into either a lepton or an antilepton and an off-shell $W$ boson, which in turn decays into two jets. If the neutrino superposition decays into a lepton and an off-shell $W$ boson,  the process is lepton number conserving (LNC) and if it decays into an antilepton it is lepton number violating (LNV). Although we focus on this example process, our results can be readily adapted to other processes as well.\footnote{The process in which the initial $W^{+}$ is replaced by a $W^-$ differs only in the leptonic mixing factors (later denoted by $V$), which are complex conjugate to the ones appearing in the process with an initial $W^+$.} 

\noindent Starting with the general formula for the connected amplitude
\begin{equation}
\mathcal{A} = \bra{f} \hat{T} \bigg( \exp\left( -i \int \mathrm{d}^4x ~ \mathcal{H}_I \right) \bigg) - \mathbf{1} \ket{i}~,
\end{equation}
where $\mathcal{H}_I$ is the interaction Hamiltonian and $\hat{T}$ the time ordering operator, we follow the procedure described in \cite{Beuthe:2001rc} for the LNC and LNV processes separately.
\\\\
Using standard QFT methods in the canonical quantisation formalism, the process described by the Feynman diagrams \subref{feyn:LNC} and \subref{feyn:LNV} is obtained by expanding the exponential up to second order in the electroweak coupling constant. The external states are considered to be wave packets that are centred at the space-time points of production or detection as in \cite{Beuthe:2001rc}. This introduces integrals over the momenta of the external particles together with functions describing the shape of the wave packets, which we denote as $\Psi$, as well as space-time translation operators in the form of exponentials accompanying each wave packet. The propagator of the intermediate particle is obtained by contraction of the relevant fields in configuration space. Using the Fourier transformation it is written down in momentum space, introducing the integral over the neutrino momentum $p$. The relevant Feynman rules, which are applicable also to the LNV process, can be found in \cite{Denner:1992vza}. The amplitudes for the processes described by the Feynman diagrams \subref{feyn:LNC} and \subref{feyn:LNV} can be written as
\begin{equation}
\mathcal{A}_{\alpha \beta}^{LNC} := \sum_j V_{j \beta} \mathcal{A}_j^{LNC} V^{*}_{j \alpha} ~,
\end{equation}
\begin{equation}
\mathcal{A}_{\alpha \beta}^{LNV} := \sum_j V^{*}_{j \beta} \mathcal{A}_j^{LNV}  V^{*}_{j \alpha} ~,
\end{equation}
where $\alpha$ and $\beta$ denote the flavour indices of the charged leptons at production and detection and LNC and LNV refers to the lepton number violating or lepton number conserving process, respectively. The sum runs over the propagating mass eigenstates, which are denoted by the indices $i$ and $j$ in this paper. The relevant part of the lepton mixing matrix, rotating the heavy neutrino mass eigenstates into the active neutrino interaction eigenstates is denoted by $V$ and the so-called partial amplitude $\mathcal{A}_j^{LNX}$ is defined as
\begin{equation}
\label{eq:partial_amplitude}
\begin{split}
\mathcal{A}_j^{LNX} :=& 
\int \mathrm{d^4 x} \int \mathrm{d^4 x'}
\int [\mathrm{d\mathbf{q}}] \Psi(\mathbf{q},\mathbf{Q}) e^{-i q x}
\int [\mathrm{d\mathbf{p_1}}] \Psi^*(\mathbf{p_1},\mathbf{P_1}) e^{i p_1 x}\\ 
&\int [\mathrm{d\mathbf{p_2}}] \Psi^*(\mathbf{p_2},\mathbf{P_2}) e^{i p_2 x'}
\int [\mathrm{d\mathbf{k_1}}] \Psi^*(\mathbf{k_1},\mathbf{K_1}) e^{i k_1 x'}
\int [\mathrm{d\mathbf{k_2}}] \Psi^*(\mathbf{k_2},\mathbf{K_2}) e^{i k_2 x'}\\
&\int \frac{\mathrm{d^4p}}{(2 \pi)^4} 
e^{i p x} e^{-i p x'} ~  e^{-i p (x_1 - x_0)} M^{LNX}_j(p, Q, P_1, P_2, K_1, K_2) \frac{1}{p^2 - m_{\nu_j}^2}~,
\end{split}
\end{equation}
where $\Psi(\mathbf{k},\mathbf{K})$ describes a wave packet centred at three momentum $\mathbf{K}$. The integration measure for the three momenta is written in a short hand notation where 
$$[d\mathbf{k}] = \frac{d^3 \mathbf{k}}{(2 \pi)^3 \sqrt{2 E_k}~}\:,$$
with $E_k$ being the energy of the respective particle. The interaction amplitude $M_j^{LNX}$ is defined as the matrix element of the LNX process without the lepton mixing matrix elements and without the denominator of the propagator, where LNX can refer either to LNC or LNV. Note that we have suppressed the spin and polarisation labels of the external particles to simplify notation. The production ($x_0$) and detection ($x_1$) points in spacetime are defined with respect to the laboratory frame. Therefore the propagation distance, which is defined as $\mathbf{L}:= \mathbf{x}_1 - \mathbf{x}_0$, is also to be understood in the laboratory frame. The propagation time is given by $T:= x_1^0 - x_0^0$.
\\\\
The authors of \cite{Beuthe:2001rc} proceed with the partial amplitude by evaluating the integrals over the three momenta of the external particles, which can be done analytically if the wave packets are assumed to have a Gaussian shape and if the interaction amplitude is approximated at the mean momenta of the external particles. The approximated interaction amplitude is written as $M^{LNX}_j(p, Q, P_1, X^N) := M^{LNX}_j(p, Q, P_1, P_2, K_1, K_2)$, where $X^N$ denotes the mean momenta of the decay products of the heavy neutrinos. This is followed by the integration over $x$ and $x'$ corresponding to the production and detection vertices.
\\\\
The last step in the computation of the partial amplitude consists in the integration over the four momentum of the intermediate particle, which shows up in the propagator. The energy integral is done by the use of the Jacob-Sachs theorem \cite{Jacob:1961MASSAL}, which basically puts the intermediate particle on-shell. The integration over the three momentum is done in three regimes separately depending on the propagation distance of the intermediate particle. The following steps are valid in the longitudinal dispersion regime, which is applicable when the propagation distance is larger than the dispersion length (cf.\ \cite{Beuthe:2001rc}). In \cref{subsec:observability_conditions} it is argued that this regime is indeed the relevant one when we estimate the widths of the wave packets from measurement uncertainties. In this regime the three-momentum of the $j$-th neutrino mass eigenstate is approximated around the point of stationary phase $\mathbf{p}_{cl,j}$, which is the ``classical'' momentum given by
\begin{equation}
\label{eq:def_classical_momentum}
\mathbf{p}_{cl,j} =  m_j \gamma_{cl} \mathbf{v}_{cl} ~,
\end{equation}
where the classical velocity is defined as $\mathbf{v}_{cl} = \mathbf{L}/T$ with the corresponding gamma factor $\gamma_{cl} = (1 - |\mathbf{v}_{cl}|^2)^{-1/2}$. The classical energy is defined accordingly as
\begin{equation}
    E_{cl,j} = \sqrt{m_j^2 + |\mathbf{p}_{cl,j}|^2}~.
\end{equation}
In the LNC case, the interaction amplitude $M^{LNX}_j$ is dependent on $\mathbf{p}$, which we approximate at the point $\mathbf{p}_{cl,j}$ as well.
\\\\
The ``oscillation probability densities'' are proportional to the absolute value squared of the respective amplitudes, averaged over the macroscopic propagation time
\begin{equation}P^{LNX}_{\alpha \beta}(\mathbf{L}, Q, P_1, X^N) \propto \int dT \mathcal{A}_{\alpha \beta}^{LNX} (\mathcal{A}_{\alpha \beta}^{LNX})^*~.
\end{equation}
To proceed with this integration Laplace`s method is used in which the macroscopic propagation time is expanded around $\tilde{T}_0$ (cf.~\cite{Beuthe:2001rc})
\begin{equation}
\label{eq:T_approx_Ttilde}
    T \approx \tilde{T}_0 = \frac{\tilde{E}_0 |\mathbf{L}|}{|\mathpzc{p}_0|}~,
\end{equation}
where $\mathpzc{p}_0 := \mathbf{p}_0 \cdot \mathbf{L}/|\mathbf{L}|$. The mean energy is defined as $\tilde{E}_0 := \sqrt{\tilde{m}_0^2 + |\mathpzc{p}_0|^2}$, with the arithmetic mean of the heavy neutrino masses $\tilde{m}_0$. The four momentum $p_0$ is defined via the mean momenta of the external particles using energy-momentum conservation at the production and/or detection vertex, which yields
\begin{equation}
\label{eq:definition_p0}
    p_0 := Q - P_1 = P_2 + K_1 + K_2~.
\end{equation}
A related mass can be defined as
\begin{equation}
m_0^2 := E_0^2 - \mathbf{p}_0^2~.
\end{equation}
Note that these so called ``reconstructed'' quantities, which are labelled with an subscript $(~)_0$, represent experimentally accessible quantities. As described below (see \cref{subsec:observability_conditions}), the reconstructed mass $m_0$ is related to the physical masses of the heavy neutrinos. In particular we note that if the heavy neutrinos are almost mass degenerate such that $m_i \approx \tilde{m}_0$, \cref{eq:m0_observability_condition} can be used to argue that $\tilde{m}_0 \approx m_0$ within the momentum uncertainty given by the wave packets widths. Using \cref{eq:T_approx_Ttilde} yields
\begin{equation}
    \mathbf{p}_{cl,j} = m_j \frac{1}{\sqrt{1- \frac{|\mathbf{p}_0|^2}{\tilde{E}_0^2}}} \frac{|\mathpzc{p}_0|}{\tilde{E}_0} \frac{\mathbf{L}}{|\mathbf{L}|}~.
\end{equation}

\noindent Due to the wave packet nature of the intermediate particle there is a non-zero probability to measure the decay vertex in a direction $\mathbf{L}$ from the production vertex not parallel to $\mathbf{p}_0$. However, those orthogonal directions are exponentially suppressed and negligible if the momentum uncertainties are small, i.e.\ if $\sigma_p$ (see \cref{apx:formulas_observability_conditions}) is smaller than the orthogonal momentum $\mathbf{p}_0 \times \mathbf{L}/|\mathbf{L}|$. We therefore introduce an integration over the direction of $\mathbf{L}$, which is evaluated by approximating $\mathbf{L}/|\mathbf{L}| \approx \mathbf{p}_0/|\mathbf{p}_0|$. This approximation allows to identify $\mathpzc{p}_0 = |\mathbf{p}_0|$. Also, with $m_j \approx m_0$, which holds for nearly mass degenerate heavy neutrinos as mentioned above, the four momenta $p_{cl,i}$ and $p_0$ are approximately equal.
\\\\
The time integration together with the integration over the direction of $\mathbf{L}$ leads to
\begin{equation}
\begin{split}
\label{eq:probability_density_LNX_general}
    &P^{LNX}_{\alpha \beta}(L,Q,P_1,X^N) := \int_{4 \pi} L^2 d\Omega_\mathbf{L} P^{LNX}_{\alpha \beta}(\mathbf{L},Q,P_1,X^N) \\
    &\quad\phantom{:}= \sum_{i,j} N_g^2 M^{LNX}_i(p_{cl,i}, Q,P_1,X^N)(M^{LNX}_j(p_{cl,j}, Q,P_1,X^N))^*\\
    &\quad\qquad \times \mathcal{V}^{LNX}_{\alpha \beta~ ij} \exp\left( -2 \pi i \frac{L}{L_{ij}^{osc}} \right) ~,
\end{split}
\end{equation}
where $L := |\mathbf{L}|$ has been defined. 
We note that in the case of the no-dispersion regime, following the steps in \cite{Beuthe:2001rc}, one obtains the same equation \cref{eq:probability_density_LNX_general} and therefore the next considerations (until \cref{subsec:observability_conditions}) hold for both regimes. The proportionality constant $N_g^2$ can be obtained by the following normalisation condition, which has to be computed $\forall \{L,\alpha\}$
\begin{equation}
\label{eq:normalisation_condition_general}
\sum_{\text{spins}} \int_{\text{PS}^N} dX^N ~\sum_\beta \bigg( \left[ P^{LNC}_{\alpha \beta}+ P^{LNV}_{\alpha \beta}\right](L,Q,P_1,X^N) \bigg) = 1~,
\end{equation}
where $\int dX^N$ denotes an integral over the whole phase space of the decay products of the heavy neutrino ${\text{PS}^N}$ and $\sum_{\text{spins}}$ represents the sum over all outgoing spins/polarisations and the average over all incoming spins/polarisations. Lepton mixing matrix factors are contained in $\mathcal{V}^{LNX}_{\alpha \beta ~ ij} $, which is defined as
\begin{equation}
\begin{split}
\mathcal{V}^{LNC}_{\alpha \beta ~ ij} &:= V_{\beta i}  V^{*}_{\alpha i} V^{*}_{\beta j}  V_{\alpha j}~,\\
\mathcal{V}^{LNV}_{\alpha \beta ~ ij} &:= V^{*}_{\beta i}  V^{*}_{\alpha i} V_{\beta j}  V_{\alpha j}~.
\end{split}
\end{equation}
The oscillation length is given by 
\begin{equation}
L_{ij}^{osc} = \frac{4 \pi |\mathbf{p}_0|}{m_{i}^2-m_{j}^2}~,
\end{equation}
where $\mathbf{p}_0$ is defined in \cref{eq:definition_p0}. Additional terms which can be neglected, given the adequate kinematic and experimental conditions, are discussed in \cref{subsec:observability_conditions}.
\\\\
\sloppy Together with the normalisation condition, the oscillation probability densities $P^{LNX}_{\alpha \beta}(L, Q, P_1, X^N)$ are defined as densities with respect to the mean momenta of the decay products of the heavy neutrino. If this density is integrated over the considered phase space of the heavy neutrino decay products $PS^N_p \subset PS^N$ an oscillation probability is obtained as
\begin{equation}
\label{eq:probability_LNX_general}
   P^{LNX}_{\alpha \beta}(L, Q, P_1) = \int_{\text{PS}^N_p} dX^N P^{LNX}_{\alpha \beta}(L, Q, P_1, X^N)~.
\end{equation}
The results describe the probabilities that the superposition of heavy neutrino mass eigenstates, produced by the decay of a $W$ boson together with an antilepton of flavour $\alpha$, produces an (anti)lepton of flavour $\beta$ if it decays after a distance $L$ in the direction of $\mathbf{p}_0$ via an LNC (LNV) process.\footnote{The normalisation condition is such that decays into other final states than the ones given in \subref{feyn:LNC} and \subref{feyn:LNV} are not considered.} When summing these probabilities over the flavour of the final (anti)lepton, the resulting quantity $\sum_\beta P^{LNV}_{\alpha \beta}(L, Q, P_1)$ can be interpreted as the probability that the produced heavy neutrino interaction eigenstate has oscillated into a heavy antineutrino interaction eigenstate. The quantity $\sum_\beta P^{LNC}_{\alpha \beta}(L, Q, P_1)$ can be interpreted as the probability that the heavy neutrino interaction eigenstate has ``survived'', i.e.\ has not oscillated into a heavy antineutrino interaction eigenstate.
\\\\
At this point \cref{eq:probability_LNX_general} together with the additional terms, to be discussed in \cref{subsec:observability_conditions}, can be regarded as the most general result. We now proceed simplifying it, in order to gain further insight. To this end, we first show under which conditions the interaction amplitudes can be factored out of the sum over mass eigenstates, and subsequently be absorbed into the normalisation constant.\footnote{We note that in the case of light neutrino flavour oscillations, when the possibility of oscillating into light antineutrinos is neglected, this is always possible. On the contrary, in the here considered case of heavy neutrino-antineutrino oscillations it is an approximation.} After that we commit to a specific model within the SPSS, which features two almost degenerate heavy neutrinos.
\\\\
The interaction amplitude is dependent on the masses of the propagating neutrinos through the numerator of the propagator, which reads $(\slashed{p}_{cl,j} + m_j)$. The neutrino masses are expressed as a deviation from the mean neutrino mass $\tilde{m}_0$.\footnote{The following reparameterization can easily be extended to more than two mass eigenstates.} In the case of just two heavy neutrinos, with masses $m_4$ and $m_5$, where w.l.o.g. $m_4 < m_5$ can be assumed, one finds that
\begin{equation}
\begin{split}
\label{eq:mass_splitting}
m_4 &= (1 - \lambda_m) \tilde{m}_0 \\
m_5 &= (1 + \lambda_m) \tilde{m}_0 ~,
\end{split}
\end{equation}
where the dimensionless mass splitting parameter $\lambda_m$ is defined as
\begin{equation}
\label{eq:def_masssplitting_parameter}
 \lambda_m = \frac{m_5 - m_4}{m_5 + m_4}~,
\end{equation}
and the mean mass is just given by
\begin{equation}
\label{eq:def_mean_mass}
    \tilde{m}_0 = (m_4 + m_5)/2~.
\end{equation}
Using the definition \cref{eq:def_classical_momentum}, the classical momentum can also be expressed using the mass splitting parameter, which yields
\begin{equation}
\begin{split}
\label{eq:momentum_splitting}
\mathbf{p}_{cl,4} &= (1 - \lambda_m) \tilde{\mathbf{p}}_{cl,0} \\
\mathbf{p}_{cl,5} &= (1 + \lambda_m) \tilde{\mathbf{p}}_{cl,0} ~,
\end{split}
\end{equation}
where the mean momentum is defined as $ \tilde{\mathbf{p}}_{cl,0} =  \tilde{m}_0 \gamma_{cl} \mathbf{v}_{cl} $. Note that using \cref{eq:mass_splitting,eq:momentum_splitting} it is easy to reparameterize the four momentum of an on-shell particle. This makes it possible to factor out the mass dependence from the interaction amplitudes, yielding
\begin{equation}
M^{LNX}_i(p_{cl,i}, Q,P_1,X^N)(M^{LNX}_j(p_{cl,j}, Q,P_1,X^N))^* = \Lambda_{ij} |M^{LNX}(\tilde{p}_{cl,0},Q,P_1,X^N)|^2~,
\end{equation}
where $\Lambda_{ij}$ contains the factors describing the mass and momentum splitting and is given by
\begin{equation}
 \Lambda_{ij} =\begin{cases}
                                   (1-\lambda_m)^2 & \text{if $i = j = 4$} \\
                                   (1+\lambda_m)^2 & \text{if $i = j = 5$} \\
  							    (1-\lambda_m^2) & \text{if $i \neq j$}~. 
  \end{cases}
\end{equation}
\\\\
The interaction amplitudes of the processes \subref{feyn:LNC} and \subref{feyn:LNV} can be written down by using the Feynman rule conventions described in \cite{Denner:1992vza}, which are applicable also to the lepton number violating diagram. This yields 
\begin{equation}
\begin{split}
\label{eq:LNC_matrixelement}
i M_j^{LNC}(p_{cl,j}, Q,P_1,X^N) &= i \frac{4 G_F}{\sqrt{2}}\bigg(\overline{u}(K_1) ~ \Gamma^\nu ~ v(K_2)\bigg)\\&\qquad\bigg(\overline{u}(P_2) ~ \Gamma_\nu ~ i (\slashed{p}_{cl,j} + m_j) ~ \frac{i g_2}{\sqrt{2}} ~ \Gamma_\mu ~ v(P_1)\bigg)  \epsilon^*_\mu(Q)~,
\end{split}
\end{equation}
and
\begin{equation}
\begin{split}
\label{eq:LNV_matrixelement}
i M_j^{LNV}(p_{cl,j}, Q,P_1,X^N) &=i \frac{4 G_F}{\sqrt{2}} \bigg(\overline{u}(K_2) ~ \Gamma^\nu ~ v(K_1)\bigg)\\&\qquad\bigg(\overline{u}(P_2) ~ \Gamma'_\nu ~ i (\slashed{p}_{cl,j} + m_j) ~ \frac{i g_2}{\sqrt{2}} ~ \Gamma_\mu ~ v(P_1)\bigg)  \epsilon^*_\mu(Q)~,
\end{split}
\end{equation}
where $g_2$ is the coupling of the $SU(2)$ gauge bosons and $G_F$ is the Fermi constant. The vertices $\Gamma^\mu$ and $\Gamma'^\mu$ are given by $\gamma^\mu ~ P_L$ and $-\gamma^\mu ~ P_R$ respectively, with the left and right chirality projection operators defined as
\begin{equation}
    P_{L/R} := \frac{1}{2}(1 \mp \gamma^5)~.
\end{equation}
Note that, as above we have suppressed spinor and color indices as well as indices denoting the spin and polarisation of the external particles. For the neutrino we chose the fermion flow from left to right in \subref{feyn:LNC} and \subref{feyn:LNV}. 
\\\\
Further simplifications are possible if the spin correlation between the production and detection vertex are neglected, i.e.\ if the numerator of the propagator can be written as 
\begin{equation}(\slashed{\tilde{p}}_{cl,0} + \tilde{m}_0) = \sum_s \overline{u}_s(\tilde{p}_{cl,0})u_s(\tilde{p}_{cl,0}) \approx   \sum_{s,s'} \overline{u}_s(\tilde{p}_{cl,0})u_{s'}(\tilde{p}_{cl,0})~.
\end{equation}
This approximation is also done in the narrow width approximation and makes it possible to factorize the interaction amplitude into a production interaction amplitude and a detection interaction amplitude. Using this approximation, the interaction amplitudes for the LNC and LNV process are identical.
\\\\
If the spin correlation is not neglected, the interaction amplitudes of the LNC and LNV process differ due to the chirality structure. For a given process it could be possible that the interaction amplitudes depend on the orientation of the momenta of the external particles in such a way that a probabilistic classification into LNC or LNV becomes possible, which could be interesting, e.g.\ for the SHiP experiment \cite{Tastet:2019nqj}.
\\\\
In order to simplify the expression \cref{eq:probability_density_LNX_general}, following the above discussion, the spin correlation between the production and detection vertices are neglected. This allows to absorb the mass splitting independent parts of the interaction amplitudes in \cref{eq:probability_density_LNX_general} into the normalisation constant. 
\\\\
This leads to the following oscillation probability, which is independent of the mean momenta of the decay products of the heavy neutrino, spins and polarisations of the external particles
\begin{equation}
P^{LNX}_{\alpha \beta} (L,Q,P_1) := \sum_{\text{spins}}\int_{\text{PS}^N} dX^N~P^{LNX}_{\alpha \beta} (L, Q,P_1,X^N)~.
\end{equation}
Due to these simplifications the oscillation probability only depends on $Q$ and $P_1$ in the combination $|\mathbf{p}_0| = |\mathbf{Q} - \mathbf{P}_1|$, which yields
\begin{equation}
\label{eq:probability_LNX}
P^{LNX}_{\alpha \beta} (L,|\mathbf{p}_0|) = \sum_{i,j} N^2 \Lambda_{ij} ~\mathcal{V}^{LNX}_{\alpha \beta ~ ij}  \exp\left( -2 \pi i \frac{L}{L_{ij}^{osc}} \right)~.
\end{equation}
The normalisation condition for the simplified oscillation probability is given by
\begin{equation}
\label{eq:normalisation_condition}
\sum_\beta \bigg( \left[P^{LNC}_{\alpha \beta} + P^{LNV}_{\alpha \beta}\right] (L,|\mathbf{p}_0|) \bigg) = 1\qquad\forall \{ L,|\mathbf{p}_0|\}~.
\end{equation}
In \cref{sec:approximations_oscillation_formula} the normalisation constant is evaluated explicitly for a specific example model of interest.
\\\\
To proceed further we assume the experimental conditions and model parameters to be such that the heavy neutrinos travel a macroscopic distance before they decay, forming a displaced vertex. The number of expected events that feature such a displaced vertex can be expressed as a formula similar to the one described in \cite{Antusch:2017hhu}. This formula has to be modified in order to cover the circumstances of this paper. In particular an expression for the probability that the heavy neutrino decays in an LNC (LNV) manner involving a specific lepton flavour between a minimum and maximum distance is needed. Following the discussion in \cite{Beuthe:2001rc} regarding unstable oscillating particles, it is shown that the only relevant modification to the oscillation formula is given by the exponential discussed in \cref{subsec:observability_conditions}. If the masses and decay widths of the heavy neutrinos are nearly identical, it is possible to define a common decay length as
\begin{equation}
\tilde{L}^{decay}_0 := \frac{p_0}{\tilde{m}_0 \tilde{\Gamma}_0}~,
\end{equation} 
where the common decay width is defined as
\begin{equation}
\label{eq:approx_decay_width}
\tilde{\Gamma}_0:= \frac{\Gamma_4 + \Gamma_5}{2}~.
\end{equation}
The exponential
\begin{equation}
    \exp \left( - \frac{L}{\tilde{L}^{decay}_0}  \right)
\end{equation}
describes the probability that a particle is still present at distance $L$. Therefore the probability density that the particle decays at distance $L$ is given by the derivative
\begin{equation}
 -\frac{d}{dL} \exp \left( - \frac{L}{\tilde{L}^{decay}_0}  \right)~.
\end{equation}
With this the probability that a particle decays in an LNX process into flavour $\beta$ between $x_{min}(\vartheta)$ and $x_{max}(\vartheta)$ is given by
\begin{equation}
\label{eq:probability_displaced_vertex}
P_{dv~\alpha\beta}^{LNX}(x_{min}(\vartheta),x_{max}(\vartheta),|\mathbf{p}_0|) = \int_{x_{min}(\vartheta)}^{x_{max}(\vartheta)} P^{LNX}_{\alpha \beta} (L,|\mathbf{p}_0|)  \bigg( -\frac{d}{dL} \exp\left(  - \frac{L}{\tilde{L}^{decay}_{0}}\right)\bigg) dL ~,
\end{equation}
where the subscript $dv$ stands for displaced vertex and $\vartheta$ denotes the angle of the heavy neutrino with respect to the beam axis. Usually the interval $[x_{min}(\vartheta),x_{max}(\vartheta)]$ will be chosen to lie inside the detector, such that the decay products can be measured. The detector geometry can be taken into account by the dependence on $\vartheta$. The number of expected LNX events in which an antilepton of flavour $\alpha$ is measured at the production vertex and a lepton (LNC) or antilepton (LNV) of flavour $\beta$ is measured at the detection vertex is given by
\begin{equation}
\label{eq:number_of_events}
N^{LNX}_{\alpha \beta} = \tilde{\sigma}_{N,0} ~ \tilde{Br}_{l j j,0} ~\mathcal{L} ~ \int D_N(\vartheta, |\mathbf{p}_0|)~ P_{dv~\alpha\beta}^{LNX}(x_{min}(\vartheta),x_{max}(\vartheta),|\mathbf{p}_0|)~d\vartheta d|\mathbf{p}_0|~,
\end{equation}
where $\tilde{\sigma}_{N,0}$ is the mean production cross section of the heavy neutrinos, which depends on model parameters such as the masses of the heavy neutrinos and the details of the lepton mixing matrix. $\tilde{Br}_{l j j,0}$ is the mean branching ratio for the decay of a heavy neutrino into a lepton and two jets and $\mathcal{L}$ is the time integrated luminosity. Therefore, the factor $\tilde{\sigma}_{N,0} ~ \tilde{Br}_{l j j,0} ~\mathcal{L}$ describes the total number of events in which a heavy neutrino is produced and decays into a lepton and two jets. It is assumed that the branching ratios as well as the production cross sections for different mass eigenstates are nearly identical in order for such an approximation to be appropriate. The remaining factor $D_N(\vartheta, |\mathbf{p}_0|)$ accounts for the probability density that the reconstructed heavy neutrino has momentum of modulus $|\mathbf{p}_0|$ and is produced with an angle $\vartheta$ with respect to the beam axis. As discussed above, $P_{dv~\alpha\beta}^{LNX}(x_{min}(\vartheta),x_{max}(\vartheta),|\mathbf{p}_0|)$ gives the probability that the heavy neutrino decays in an LNX manner into flavour $\beta$ inside the interval $[x_{min}(\vartheta),x_{max}(\vartheta)]$. Note that if one wanted to consider the spin correlation, one has to use \cref{eq:probability_LNX_general} in the definition of \cref{eq:probability_displaced_vertex}.

\subsection{Observability Conditions and Dispersion Length}
\label{subsec:observability_conditions}
In order for the oscillations to be observable there are several conditions which have to be satisfied. This subsection describes those conditions and estimates their viability for typical parameters of long-lived heavy neutrinos detectable at, e.g.\ HL-LHCb. We consider as an explicit example a parameter point for the minimal low scale linear seesaw model that has also been used in Ref.~\cite{Antusch:2017ebe}.
\\\\
To compute all relevant parameters it is necessary to know the kinematics of the process and the widths of the external wave packets. One approach to estimate these widths is based on the following considerations: Since the final particles at production and detection are reconstructed by measurements at a detector, the uncertainty of the measurement should be reflected by the widths of the respective wave packets. For charged leptons a relative momentum uncertainty in the range $(\Delta p / p)_{lepton} = [0.5\%, 1\%]$ holds for particles with a long enough track, cf.~\cite{Aaij:2014jba}, which we therefore use for the widths of the charged lepton wave packets. The momentum resolution for the quarks, which are reconstructed from displaced jets, is much harder to determine. For a conservative estimate we therefore use a large range for their relative momentum uncertainty $(\Delta p / p)_{quark} = [5\%, 30\%]$ (and thus for the possible widths of the quark wave packets). Finally, the width of the wave packet of the initial $W$ boson is taken to be its decay width ($\Gamma_W \approx 2 \: \mbox{GeV}$). 
\\\\
Alternatively, one can also try to estimate the widths of the wave packets in position space based on the consideration that the uncertainty is determined by interactions with detector/beam particles and their respective widths. For the $W$ boson one could use the proton-proton distance in the proton bunches, which at the LHC is about $5 \times 10^{-6}$ cm, whereas for the leptons and quarks one might take a wave packet width of the order of an atom radius, i.e.\ about $10^{-8}$ cm. This would lead to significantly smaller widths in momentum space compared to the estimate using the measurement uncertainty, such that the appropriate regime is the no-dispersion regime (cf.\ \cite{Beuthe:2001rc}). We have checked that the relevant observability conditions in this case are all satisfied for our example parameter point, and our results from \cref{eq:probability_LNX_general,eq:number_of_events,eq:LNC_probability_expansion,eq:LNV_probability_expansion} can also be used for these estimates of the wave packet widths. From now on we will focus on the estimates for the wave packet widths from the momentum space considerations.    
\\\\
The kinematics of the processes described by \subref{feyn:LNC} and \subref{feyn:LNV} have been simulated, assuming two nearly degenerate heavy neutrinos. Since the purpose of the simulation is to compute the parameters necessary to check the observability conditions and not to simulate the oscillation process itself, all particles can be treated as plane waves, where their momentum represents the momentum of the peak of the wave packet. Simulating enough events, where the momentum of the $W$ boson is taken in the range 340 GeV to 2 TeV. It is then possible to check if the observability conditions are fulfilled. For the simulation the parameter values in \cref{table:parameter_simulation} have been used. It has also been assumed for simplicity that there are two heavy neutrino mass eigenstates with masses $m_4$ and $m_5$.
\begin{table}
\centering
 \begin{tabular}{|c c|} 
 \hline
 parameter & value \\
  \hline\hline
$\gamma$ & $\approx 50$\\
 $\tilde{m}_0 ~[\mbox{GeV}]$ & 7\\
 $\delta m_{45}^2~ [\mbox{GeV}^2]$ & $-1.04 \times 10^{-11}$ \\
  \hline
 $(\Delta p / p)_{lepton}$ &  $0.5\%$ -- $1\%$\\
 $(\Delta p / p)_{quark}$ &$5\%$ -- $30\%$\\ [0.5ex] 
 \hline
\end{tabular}
\caption{Parameters used for simulating the kinematics of the processes in \subref{feyn:LNC} and \subref{feyn:LNV}, in order to evaluate the observability conditions for the example minimal low scale seesaw parameter point used in \cite{Antusch:2017ebe}. $\gamma$ denotes the gamma factor of the heavy neutrinos, $\tilde{m}_0$ their mean mass (which in the simulation coincides with $m_0$), and the squared mass splitting $\delta m_{45}^2 = m_4^2 - m_5^2$ in the scenario of \cite{Antusch:2017ebe} is predicted by the measured values of the light neutrino mass splittings. The used range for the uncertainties in the measurement of the momenta of the external charged leptons and jets are denoted as $(\Delta p / p)_{lepton}$ and  $(\Delta p / p)_{quark}$.}
\label{table:parameter_simulation}
\end{table}
\\\\
The observability conditions are given as exponential suppression factors. If it is not clear that they are satisfied those exponential factors have to be included into the probability \cref{eq:probability_density_LNX_general} or \cref{eq:probability_LNX}, respectively. It is worth mentioning that including exponential terms into the probability changes the normalisation constant, which can be computed using \cref{eq:normalisation_condition_general} or \cref{eq:normalisation_condition}. The quantities used in the computation of the observability conditions are defined in \cref{apx:formulas_observability_conditions}.
\\\\
The \emph{oscillation length} sets the scale of the experiment, since it is the length at which the measurements should be taken in order to observe oscillations. As stated above it is given by
\begin{equation*}
L_{ij}^{osc} = \frac{4 \pi |\mathbf{p}_0|}{m_{i}^2-m_{j}^2}~.
\end{equation*}
With the parameters in \cref{table:parameter_simulation} (with $\gamma = 50$) it can be computed to be
\begin{equation}
L_{45}^{osc} \approx 8.34~ \text{cm}~,
\end{equation}
where the two heavy neutrino mass eigenstates are labelled with subscript numbers 4 and 5.
\\\\
The \emph{effective width} $\sigma_{peff}$ (see \cite{Beuthe:2001rc}) can be interpreted as the width of the wave packets of the heavy neutrinos. A rough estimate of the effective width can be obtained by neglecting the detection process and by approximating the lepton in the production process as a plane wave. Due to energy-momentum conservation, the shape of the effective wave packets of the heavy neutrinos is then given by the one of the initial $W$ boson, which width is approximated by its decay width. However, simulating the kinematics of the process and computing the width numerically shows that it is in the range
\begin{equation}
\label{eq:effective_width_range}
\sigma_{peff}^{sim} \in [4.5 \times 10^{-3},~32] ~\mbox{GeV}~,
\end{equation}
showing that the estimate $\sigma_{peff} \approx 2~\mbox{GeV}$ is indeed very rough.
\\\\
The \emph{dispersion length} is the threshold at which the spread of the wave packets of the heavy neutrinos becomes significant in all directions. At distances larger than the dispersion length the methods of the longitudinal dispersion regime have to be used. The dispersion length is given by \cite{Beuthe:2001rc}
\begin{equation}
L^{disp}_j = v_0 \frac{E_0^2}{2 m_j^2 \sigma_{peff}^2}~,
\end{equation}
where $v_0 = |\mathbf{p}_0|/E_0$ with $\mathbf{p}_0$ and $E_0$ being the momentum and the energy given by the mean momenta of the external particles and energy-momentum conservation at either vertex as defined above. The simulation using the parameter values from \cref{table:parameter_simulation} shows that $L_{osc} >1000~L_{disp}$. The assumption that the longitudinal dispersion regime is the relevant one, in the case in which the wave packet widths are estimated from measurement uncertainties, is therefore well justified.
\\\\
The observability conditions can be divided into two groups. The so-called coherent effects, which are taken into account at the level of the wave function and the so-called incoherent effects, which are included at the level of the probability. 

\subsubsection{Coherent Effects}
The so-called \emph{coherence length} describes the decoherence of the wave packets, which can have two origins. The oscillations either vanish if the wave packets become separated due to different group velocities, or if the wave packets spread beyond the oscillation length, in which case the oscillations are averaged to zero. In momentum space both of these effects are taken into account by the exponential
\begin{equation}
\exp\left( - \frac{L}{L^{coh}_{ij}} \right)~,
\end{equation}
where (see \cite{Beuthe:2001rc})
\begin{equation}
\label{eq:coherence_length}
L^{coh}_{ij} = \frac{1}{\sqrt{2} \pi} \frac{|\mathbf{p}_0|}{\sigma_{peff}} L^{osc}_{ij}~.
\end{equation} 
These terms can be neglected if the momentum of the intermediate particle is much larger than the width of its effective wave packet. Using the parameters in \cref{table:parameter_simulation}, all simulated events satisfy at least 
\begin{equation}
L^{coh}_{45} > 10 ~L^{osc}_{45}~.
\end{equation}
It is therefore justified to neglect the effects of decoherence in the first oscillation cycles for the parameter values in \cref{table:parameter_simulation}.
\\\\
We remark that the oscillation length and therefore the coherence length for oscillations including both the light and heavy neutrino mass eigenstates is smaller than $10^{-12}~m$. Therefore it is appropriate to neglect the light neutrino mass eigenstates in the oscillations.
\\\\
\emph{Localisation conditions} determine whether there is decoherence from the start. The relevant exponential suppressing the oscillations reads (see \cite{Beuthe:2001rc})
\begin{equation}
\exp \left( - \frac{(\delta m_{ij}^2)^2}{32 |\mathbf{p}_0|^2} \left( \frac{v_0^2}{\sigma_m^2} + \frac{\rho^2}{\sigma_{peff}^2}\right) \right)~,
\end{equation}
where $\delta m_{ij}^2 := m_i^2 - m_j^2$. The effective width of the propagating neutrino is given in momentum space by $\sigma_{peff}$. The parameters $\sigma_m$ and $\rho$ are determined by the widths of the external particles and their velocities (see \cite{Beuthe:2001rc}). Using the parameters from \cref{table:parameter_simulation} it follows that
\begin{equation}
\frac{(\delta m_{ij}^2)^2}{32|\mathbf{p}_0|^2} \left( \frac{v_0^2}{\sigma_m^2} + \frac{\rho^2}{\sigma_{peff}^2}\right) < 10^{-10}~
\end{equation}
is satisfied in all events. This allows to neglect the effects from localisation in the oscillation formula for parameter values as in \cref{table:parameter_simulation}.
\\\\
In the process considered in this paper the heavy neutrinos are \emph{unstable intermediate particles}. Therefore the full propagator should be used in \cref{eq:partial_amplitude}. As discussed in \cite{Beuthe:2001rc} this leads to the exponential decrease factor
\begin{equation}
\exp\left(  - \frac{L}{L^{decay}_{ij}}\right)~,
\end{equation} where the decay length is given by
\begin{equation}
L^{decay}_{ij} = \frac{2 |\mathbf{p}_0|}{m_i \Gamma_i + m_j \Gamma_j}~,
\end{equation}
that has to be included in the sum over the mass eigenstates in \cref{eq:probability_LNX}. With the definitions 
\begin{equation}
\label{eq:m0_and_gamma0}
    \tilde{m}_0 = \frac{m_5 + m_4}{2} \qquad \tilde{\Gamma}_0 = \frac{\Gamma_5 + \Gamma_4}{2} ~,
\end{equation}
and
\begin{equation}
\label{eq:delta_m_and_gamma}
    \delta m = m_5 - m_4 \qquad \delta \Gamma = \Gamma_5 - \Gamma_4 ~, 
\end{equation}
the decay exponential can be written as
\begin{equation}
    \exp\left(  - \frac{L}{L^{decay}_{ij}}\right) =  \exp\left(  - \frac{L (\tilde{m}_0 \tilde{\Gamma}_0 + \frac{1}{4}\delta m \delta \Gamma)}{|\mathbf{p}_0|}\right) \exp\left( \pm \frac{\delta_{ij}}{2} \frac{L (\delta m \tilde{\Gamma}_0 + \delta \Gamma \tilde{m}_0)}{|\mathbf{p}_0|}  \right)~,
\end{equation}
where for $i=j=4$ the plus sign and for $i=j=5$ the minus sign is used. The first exponential can be absorbed into the normalisation constant using the condition \cref{eq:normalisation_condition}, since it does not depend on the mass indices $i$ and $j$. If e.g.\ the decay widths of the mass eigenstates are too different, the second exponential can lead to a suppression of the oscillation pattern. On the other hand, the exponential is negligible if the mass eigenstates are nearly degenerate and if the decay widths are nearly equal. For the example point considered in this paper one can verify that $L (\delta m \tilde{\Gamma}_0 + \delta \Gamma \tilde{m}_0) \ll p_0$ such that this conditions is satisfied.
\\\\
Since the detection of the decay products allows to reconstruct the invariant mass of the propagating particle, a \emph{condition that relates the mass of the propagating particle with the detection uncertainty} is expected. This condition stems from the exponential
\begin{equation}
    \exp\left( -\frac{(\delta m_i^2 + \delta m_j^2)^2}{32 \sigma_m^2 E_0^2} \right)~,
\end{equation}
that appears in the derivation of the amplitude (see \cite{Beuthe:2001rc}). Here $\delta m_i^2 = m_i^2 - m_0^2$. It requires that
\begin{equation}
\label{eq:m0_observability_condition}
 \frac{| m_i^2 + m_j^2 - 2 m_0^2 |}{ E_0} < \sigma_m \:,
\end{equation}
where $m_i$ and $m_j$ are the masses of the heavy neutrinos and $m_0$ is as given above\footnote{We remark that $m_0$ should not be confused with $\tilde{m}_0$, which is the geometric mean of the heavy neutrino masses.}
\begin{equation}
m_0^2 := E_0^2 - |\mathbf{p}_0|^2~.
\end{equation}
The parameter $\sigma_m$ (see \cite{Beuthe:2001rc}) is related to the widths and velocities of the external particles. Since the detection process is described by the interaction with those external particles, their widths are in turn related to the precision of the momentum measurement. In conclusion this condition enforces the neutrinos to be either nearly degenerate in mass or highly relativistic, such that the propagating mass eigenstates are within the uncertainty of the momentum measurement. The simulation shows that $\sigma_m \in [0.0035~\mbox{GeV} ,32~\mbox{GeV}]$, such that the above condition is satisfied for the parameters in \cref{table:parameter_simulation}.
\\\\
In the processes considered in this paper (\subref{feyn:LNC} and \subref{feyn:LNV}) a $W$ boson is decaying in flight and is therefore an \emph{unstable source}. In principle one would have to treat the $W$ boson as a propagator connecting the diagrams \subref{feyn:LNC} and \subref{feyn:LNV} and the particles producing the $W$ boson. This extra propagator would however result in technical difficulties. The authors of \cite{Rich:1993wu,Grimus:1999ra,Grimus:1998uh} have used perturbation theory in a quantum mechanical model to describe neutrino oscillations and found an additional localisation condition, which suppresses oscillations if the unstable source moves a distance greater than the oscillation length during its lifetime. The unstable source has been assumed to have a mean momentum at rest in those derivations. A QFT approach to light neutrino oscillations has been used in \cite{CAMPAGNE1997135}, where a similar localisation condition to the one above has been derived. The authors of \cite{Dolgov:1999sp} considered a moving unstable source, i.e.\ a pion decaying in flight, in a QFT treatment. They obtained the constraint
\begin{equation}
    \left(2 \pi  \frac{\mathbf{v}_\pi \cdot \mathbf{v}_0}{\mathbf{v}_0^2 - \mathbf{v}_\pi \cdot \mathbf{v}_0}    \right) \frac{|\mathbf{p}_\pi|}{m_\pi \Gamma_\pi} \ll L^{osc}_{ij}~,
\end{equation}
where $\mathbf{v}_\pi, |\mathbf{p}_\pi|, m_\pi, \Gamma_\pi$ are the velocity, momentum, mass and decay width of the decaying pion. In our case the initial W boson takes the place of the pion as the unstable source. The heavy neutrino is highly boosted such that the velocities $\mathbf{v}_W$ and $\mathbf{v}_0$ are almost parallel. Furthermore it holds that $\mathbf{v}_W \ll \mathbf{v}_0$ for the parameter space considered in this paper, which implies that
\begin{equation}
     \left(2 \pi  \frac{\mathbf{v}_\pi \cdot \mathbf{v}_0}{\mathbf{v}_0^2 - \mathbf{v}_\pi \cdot \mathbf{v}_0}    \right) < 2 \pi~.
\end{equation}
Even putting the momentum of the $W$ boson as 2 TeV, which is the maximum of the range considered in this paper, we find that
\begin{equation}
    \frac{|\mathbf{p}_W|}{m_W \Gamma_W} < 2.5 \times 10^{-15}~\mathrm{m}~.
\end{equation}
This constraint is therefore negligible for the process and parameter space considered. As an additional remark, note that the above mentioned localisation conditions describe the suppression stemming from the fact that the production vertex is not known due to the finite lifetime and movement of the unstable source. At current and considered future colliders the W boson decays promptly, due to its large decay width, which makes the possible decay region much smaller than any macroscopic oscillation length. Already this argument shows that the additional localisation condition, due to the instability of the W, should be satisfied.
\\\\
As discussed in \cite{Grimus:1998uh}, the unstable source can also lead to a loss of coherence for distances larger than the coherence length
\begin{equation}
    L^{coh}_\Gamma = -\frac{4 E_0^2}{\delta m_{ij}^2 \Gamma}~.
\end{equation}
For the parameters in \cref{table:parameter_simulation}, this additional coherence length can be neglected compared to the one given in \cref{eq:coherence_length}.

\subsubsection{Incoherent effects}
If the \emph{propagation distance is not precisely known}, which is the case if the production or detection points are measured with some uncertainty, neutrinos that have travelled different distances overlap and wash out the oscillation pattern. This effect can be described by the following exponential (see \cite{Beuthe:2001rc})
\begin{equation}
\exp \left( - 2 \pi^2 \left(\frac{\Delta L}{L^{osc}_{ij}} \right)^2 \right)~,
\end{equation}
where we have assumed that the propagation distance of the neutrinos is given by a Gaussian with width $\Delta L$. Oscillations vanish if $\Delta L \geq L^{osc}_{ij}$. In our case the oscillation length is around 8cm and therefore much bigger than the uncertainty in the resolution of the position of the primary and secondary vertex.
\\\\
In a real experiment there is a \emph{distribution of mean momenta of the external particles}, such that the reconstructed momentum of the intermediate particle, denoted by $\mathbf{p}_0$, follows a distribution as well. The effect is already included in \cref{eq:number_of_events}, where the factor $D_N(\vartheta, |\mathbf{p}_0|)$ describes the distribution of $|\mathbf{p}_0|$. A distribution of $|\mathbf{p}_0|$ leads to a washout of the oscillation pattern, since different oscillation lengths superimpose. In order to resolve the oscillation patterns it is therefore helpful \cite{Antusch:2017ebe} to plot the oscillation probability as a function of the reconstructed proper time using the following relations
\begin{equation}
    |\mathbf{p}_0| = m_0 \gamma_0 |\mathbf{v}_0| = m_0 \gamma_0 \frac{L}{T_0} = m_0 \frac{L}{\tau_0}~,
\end{equation}
where the reconstructed gamma factor is defined as $\gamma_0 = (\sqrt{1-|\mathbf{v}_0|^2})^{-1}$, the reconstructed velocity is as before $\mathbf{v_0} = \mathbf{p}_0/E_0$, the reconstructed time is given by $T_0 = L/|\mathbf{v_0}|$ and the reconstructed proper time is given by $\tau_0 = T_0/\gamma_0$. This leads to the following oscillation exponential
\begin{equation}
    \exp\left(-2 \pi i \frac{L}{L_{45}^{osc}}  \right) = \exp\left(- i \frac{\tilde{m}_0}{m_0} \delta m \tau_0 \right)~,
\end{equation}
where \cref{eq:def_mean_mass,eq:delta_m_and_gamma} have been used. Using an oscillation probability based on plane wave arguments, the above method has been demonstrated for an example parameter point (assuming $\tilde{m}_0 = m_0$) in \cite{Antusch:2017ebe}. Note that, as mentioned above, the quantities denoted by a subscript $(~)_{0}$ are the ones which are reconstructed by experimental measurements of the external particles in the process.

\section{Low Scale Seesaw with Symmetry Protection}
\label{sec:lowscaleseesaw}
To further develop and apply the above results, we consider SPSS models (cf.\ \cite{Antusch:2015mia,Antusch:2016ejd}) i.e.\ low scale seesaw models where the smallness of the light neutrino masses is protected by a slightly broken ``lepton number''-like symmetry. As a particular example we focus on the ``minimal low scale linear seesaw'' model with only two right-handed (sterile) neutrinos, that has also been discussed as an example in \cite{Antusch:2017ebe}. 
The Lagrangian of this model takes the following form
\begin{equation}
\label{eq:lagrangian}
\mathcal{L} = \mathcal{L}_{\text{SM}} - \overline{N}^1_R \Lambda (N^2_R)^c - Y_{\alpha} \overline{N}^1_R \tilde{\phi}^\dagger L^\alpha - Y'_{\alpha} \overline{N}^2_R \tilde{\phi}^\dagger L^\alpha + H.c. \:,
\end{equation}
where $\mathcal{L}_{\text{SM}} $ is the Standard Model (SM) Lagrangian, $\alpha = (e, \mu, \tau)$ is a family index and $\tilde{\phi} := \epsilon \phi^*$ with the Levi Civita symbol $\epsilon$ and the SM Higgs doublet $\phi$. The Yukawa couplings to the sterile neutrinos are denoted by $Y_\alpha$ and $Y'_\alpha$. 
In the symmetry limit of the model, the Yukawa couplings $Y'_\alpha$ are zero, and the ``lepton number''-like symmetry is only broken slightly by the Yukawa coupling $Y'_\alpha$, for which we assume that $Y'_\alpha \ll Y_\beta$ for all entries. 
Possibilities to realise a low scale linear seesaw in the context of SO(10) Grand Unified Theories have been discussed e.g.\ in \cite{Malinsky:2005bi,Antusch:2017tud}.
\\\\
After electroweak symmetry breaking the part of the Lagrangian responsible for neutrino masses and mixing can be written as
\begin{equation}
\mathcal{L}_{mass} = - \frac{1}{2} (\overline{n}^c)^T M_\nu + \text{H.c.}~,
\end{equation}
where $n = (\nu_{e_L}, \nu_{\mu_L}, \nu_{\tau_L}, (N^1_R)^c, (N^2_R)^c)^T$ and \begin{align}
 & M_\nu = 
 \begin{pmatrix}
  0 & \vec{m} &  \vec{m}'  \\
 \vec{m}^T & 0 & \Lambda\\
   (\vec{m}')^T & \Lambda & 0
 \end{pmatrix}=:\begin{pmatrix}
  0 & m  \\
 m^T & M_{\nu_h}
 \end{pmatrix}~.
\end{align}
The symbols $\vec{m}$ and $ \vec{m}'$ denote column vectors given by $\vec{Y} v_{EW} / \sqrt{2}$ and $\vec{Y}' v_{EW} / \sqrt{2}$, respectively. $v_{EW} \approx 246$ GeV is the electroweak vacuum expectation value. The mass matrix can be diagonalised using a Takagi decomposition
\begin{equation}
\label{eq_NeutrinoMassMtxTakagi}
M^D_\nu = U^T M_\nu U.
\end{equation}
This can be achieved following the steps in \cite{Antusch:2009gn} where first a block diagonalisation with an exponential ansatz is performed, followed by a diagonalisation of the active neutrino 3x3 and sterile neutrino 2x2 block. Expanding the exponential to second order yields
\begin{align}
\label{eq_FullNeutrinoMixingMatrix}
 & U = 
 \begin{pmatrix}
  1-\frac{1}{2} \theta \theta^\dagger & \theta  \\
  - \theta^\dagger & 1-\frac{1}{2} \theta \theta^\dagger
 \end{pmatrix}\begin{pmatrix}
  U_3& 0  \\
  0 & U_2
 \end{pmatrix}~,
\end{align}
where $\theta = m^* (M_{\nu_h}^*)^{-1}$. 
It can be easily checked that the mass matrix is indeed block diagonalized to second order in $\theta$.
\\\\
Furthermore, one finds that 
\begin{align}
\label{eq:HeavyNuMixingMatrix}
 & U_2 = \frac{e^{-\frac{i}{2} \arg \left( \vec{m}' ~\vec{m}^*\right)}}{\sqrt{2}}
 \begin{pmatrix}
  \frac{i  \vec{m}'~ \vec{m}^*}{\left|  \vec{m}' ~\vec{m}^*\right| } &\frac{  \vec{m}'~ \vec{m}^*}{\left|  \vec{m}' ~\vec{m}^*\right| }\\
 -i & 1 \\
 \end{pmatrix}\;
\end{align}
diagonalizes the heavy neutrino 2x2 block. 
Regarding the heavy neutrino-antineutrino oscillations, the interesting part of the mixing matrix is the upper right $3 \times 2$ block, which describes the mixing between the active neutrino interaction eigenstates and the heavy neutrino mass eigenstates. In the oscillation formula we called this part $V$. From \cref{eq_FullNeutrinoMixingMatrix,eq:HeavyNuMixingMatrix} we find that
\begin{equation}
\label{eq:mixing_matrix}
 V = \frac {1} {\sqrt{2}}
 \begin {pmatrix}
-i (\vec{\theta})^* e^{-i\phi} + i (\vec{\theta}')^* e^{i\phi} &(\vec{\theta})^*
    e^{-i\phi} + (\vec{\theta}')^* e^{i\phi}
    \end {pmatrix}\;.
\end{equation}
where the phase is defined as $ 2 \phi = \arg \left((\vec{\theta}') \cdot (\vec{\theta})^*\right)$ and the expansion parameters are defined as
\begin{align}
    \vec{\theta} &= \frac{\vec{m}}{\Lambda} = \frac{\vec{Y}~v_{EW}}{\sqrt{2} \Lambda}\\\
    \vec{\theta}' &= \frac{\vec{m}'}{\Lambda} = \frac{\vec{Y}'~v_{EW}}{\sqrt{2} \Lambda}~.
\end{align}
Note that $\theta$ can be seen as a function of $\vec{\theta}$ and $\vec{\theta}'$. From the hierarchy between the Yukawa couplings of the sterile neutrinos it follows that $\vec{\theta}' \ll \vec{\theta}$. We restrict ourselves to maximally first order in the elements of $\vec{\theta}'$ in the following.
\\\\
As a remark one can see that higher order terms in $\vec{\theta}$ can be absorbed in a slight rescaling of the Yukawa couplings $\vec{Y}$, since in the symmetry limit (where $\vec{\theta}' = 0$) $V$ has the exact form
\begin{equation}
 V_{symm} = \frac {1} {\sqrt{2 + 2|\vec{\theta}|^2}}
 \begin {pmatrix}
-i(\vec{\theta})^*  &(\vec{\theta})^*
    \end {pmatrix}\;.
\end{equation}
Therefore those terms do not qualitatively change the oscillation formulae that are obtained in \cref{sec:approximations_oscillation_formula}.
\\\\
Using the expansion \cref{eq:mixing_matrix} it is also possible to obtain expressions for the masses of the heavy neutrinos and thus also for the mass splitting parameter $\lambda_m$, introduced in \cref{eq:def_masssplitting_parameter}, which can be expressed as
\begin{equation}
\label{eq:approx_lambdam}
\lambda_m = \frac{2 |\vec{m}' \cdot \vec{m}^*|}{2 \Lambda^2 + | \vec{m}'|^2 + |\vec{m}|^2} \leq |\vec{\theta}' \cdot \vec{\theta}^*| \ll \mathcal{O}(\vec{\theta}')~.
\end{equation}
Another important parameter entering the oscillation formula is the quadratic mass splitting $\delta m_{45}^2$, which enters the formula for the oscillation length. It can be expressed as
\begin{equation}
    \delta m_{45}^2 = - 4 \tilde{m}_0^2 \lambda_m = - 2 \tilde{m}_0 \delta m~,
\end{equation}
where the mean mass of the heavy neutrinos $\tilde{m}_0$ is defined in \cref{eq:def_mean_mass} and can be expressed as
\begin{equation}
    \tilde{m}_0 =  \frac{2 \Lambda^2 + |\vec{m}|^2}{2\Lambda} + \mathcal{O}(\vec{\theta}')~.
\end{equation}

\section{Approximations of the Oscillation Formula}
\label{sec:approximations_oscillation_formula}
\begin{figure}[htbp]
    \centering
    \makebox[\textwidth]{
    \begin{minipage}{15cm}
    \begin{subfigure}[t]{0.48\textwidth}
    \centering
     \includegraphics[width=1\textwidth]{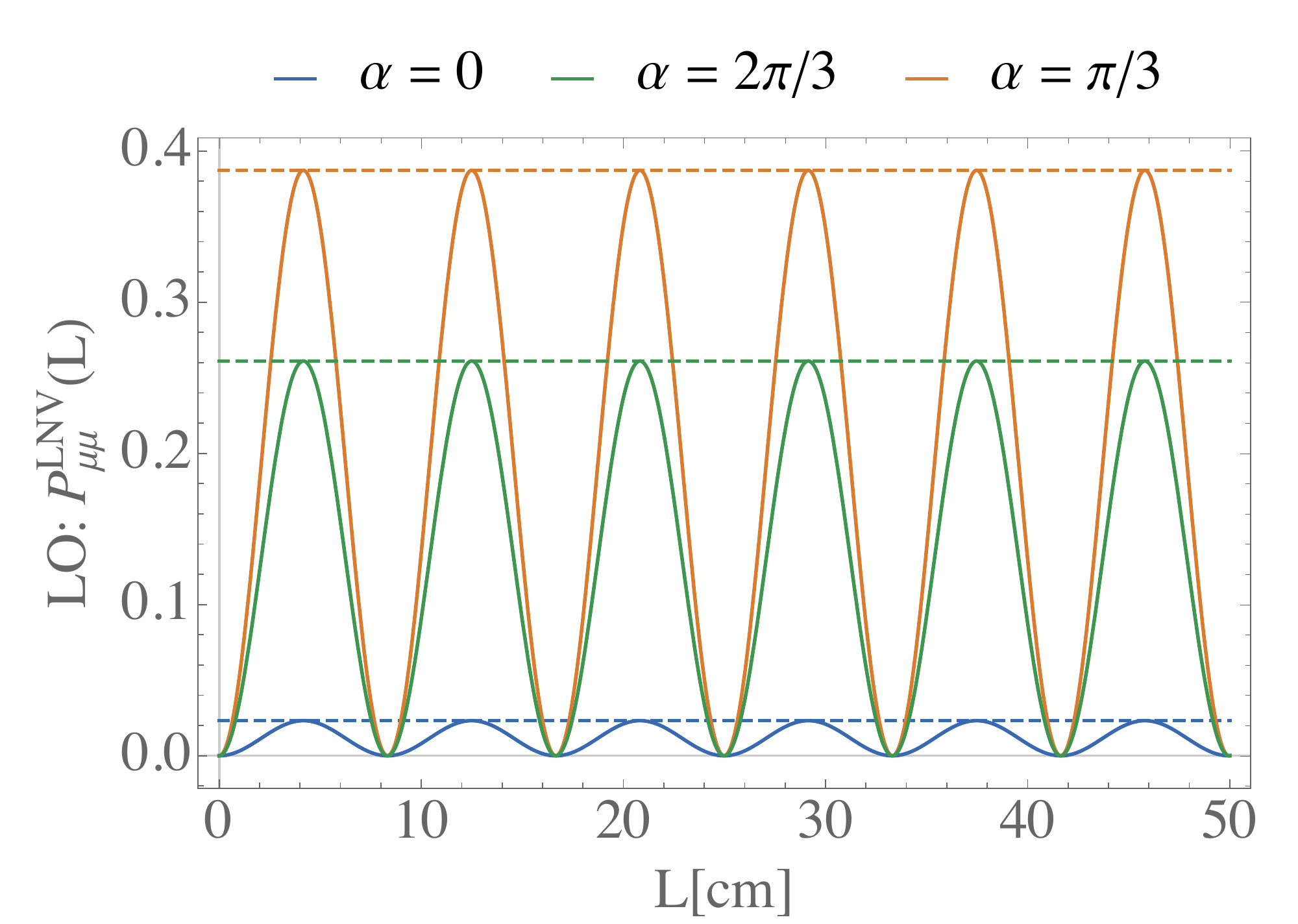}
     \caption{Heavy neutrino-antineutrino oscillation probabilities $P_{\mu \mu}^{LNV} (L)$ at LO (cf.~\cref{eq:LNV_probability_expansion}), shown as solid lines for different values of the Majorana phase $\alpha$. The dashed lines show the sum of the LNC and LNV probability ,i.e.\ $P_{\mu \mu}^{LNC} (L) + P_{\mu \mu}^{LNV} (L)$. The legend labels lines of increasing amplitude from left to right.}\label{fig:mixing_probability_LO_mumu}
    \end{subfigure}
    \hfill
    \begin{subfigure}[t]{0.48\textwidth}
    \centering
     \includegraphics[width=1\textwidth]{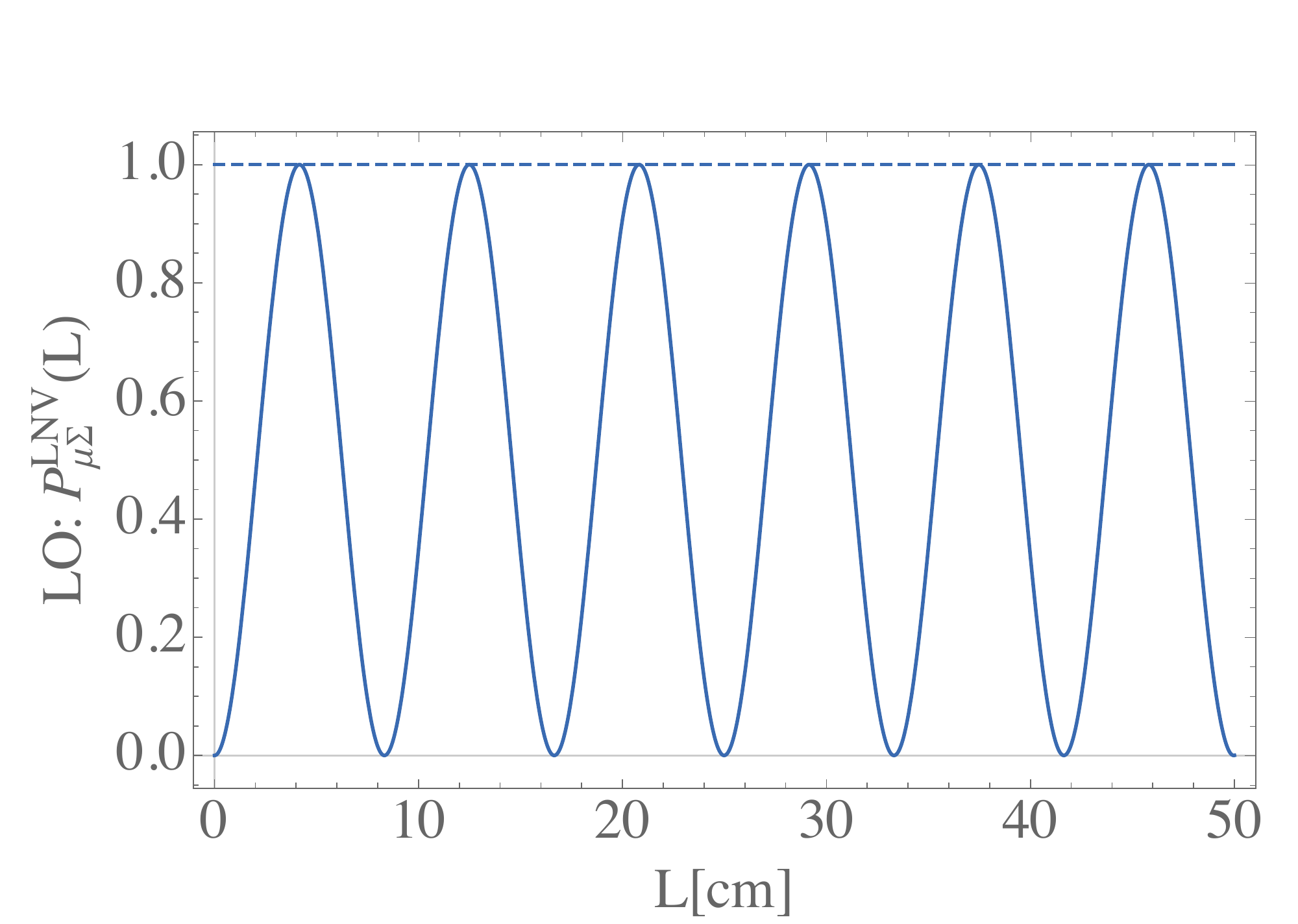}
     \caption{Heavy neutrino-antineutrino oscillation probabilities $P_{\mu \Sigma}^{LNV} (L) := \sum_\beta P_{\mu \beta}^{LNV} (L)$ at LO (cf.~\cref{eq:LNV_probability_expansion}), shown as solid lines. The dashed lines show the sum of the LNC and LNV probability ,i.e.\ $P_{\mu \Sigma}^{LNC} (L) + P_{\mu \Sigma}^{LNV} (L)$. Both probabilities are independent of the Majorana phase $\alpha$.}\label{fig:mixing_probability_LO_muSigma}
    \end{subfigure}
    \end{minipage}
    }
    \caption{Heavy neutrino-antineutrino oscillation probabilities at LO.}
\end{figure}

\begin{figure}[htbp]
    \centering
    \makebox[\textwidth]{
    \begin{minipage}{15cm}
    \begin{subfigure}[t]{0.48\textwidth}
    \centering
     \includegraphics[width=1\textwidth]{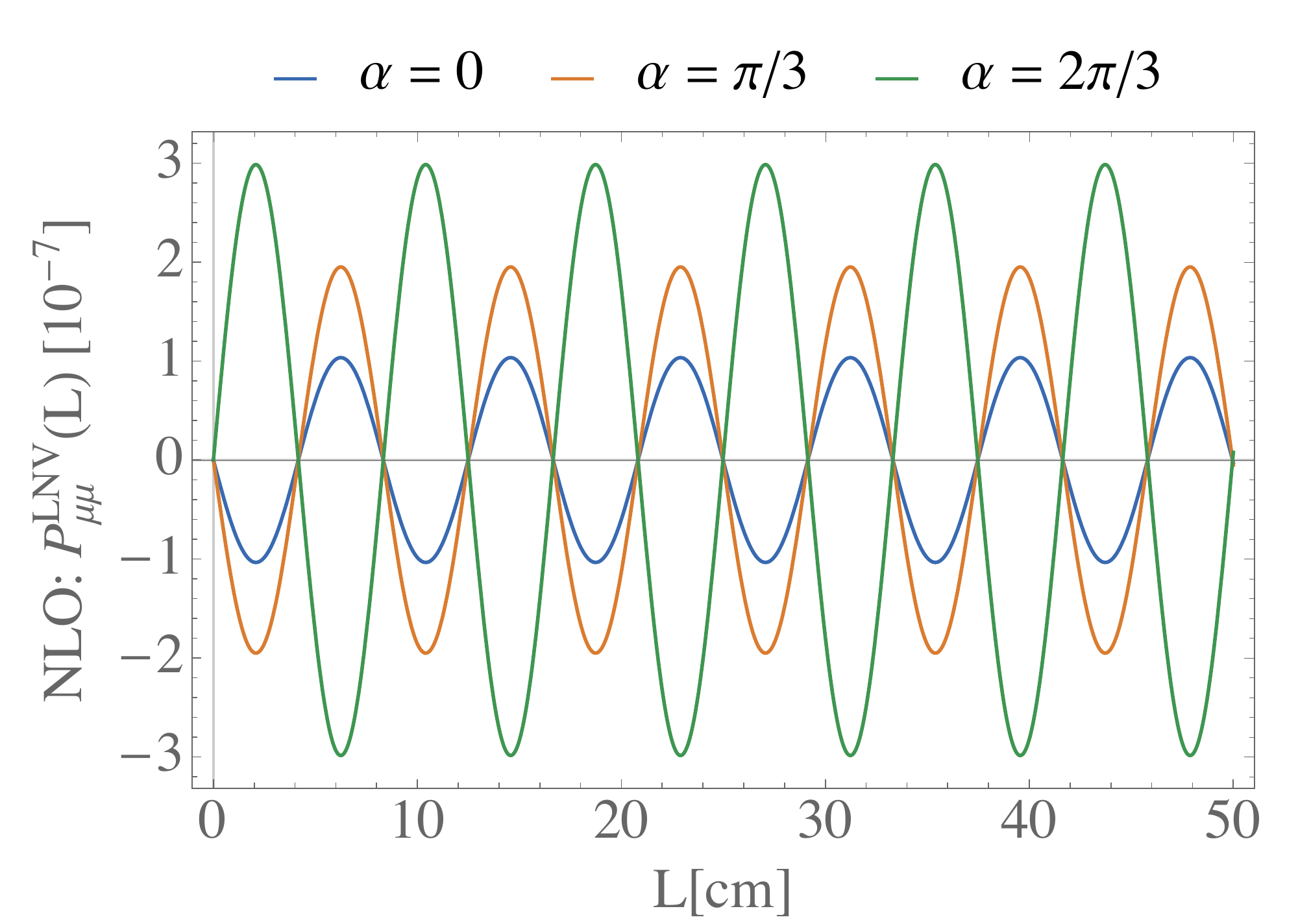}
     \caption{NLO contributions to the heavy neutrino-antineutrino oscillation probabilities $P_{\mu \mu}^{LNV} (L)$ (cf.~\cref{eq:LNV_probability_expansion}), shown as solid lines for different values of the Majorana phase $\alpha$. The NLO contribution to the LNC probability vanishes for equal incoming and outgoing flavours. The legend labels lines of increasing amplitude from left to right.}\label{fig:mixing_probability_NLO_mumu}
    \end{subfigure}
    \hfill
    \begin{subfigure}[t]{0.48\textwidth}
    \centering
     \includegraphics[width=1\textwidth]{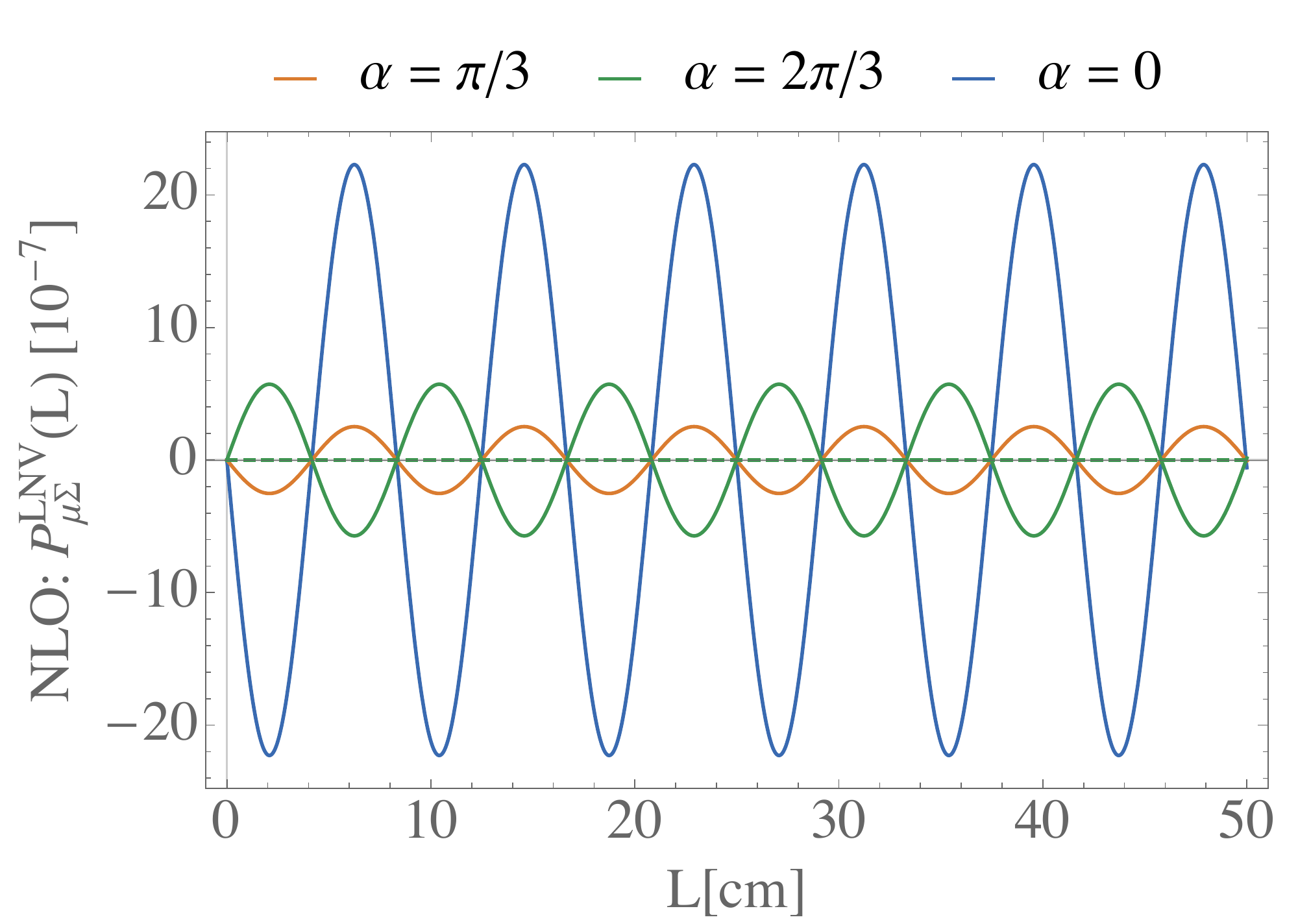}
     \caption{NLO contributions to the heavy neutrino-antineutrino oscillation probabilities $P_{\mu \Sigma}^{LNV} (L) := \sum_\beta P_{\mu \beta}^{LNV} (L)$ (cf.~\cref{eq:LNV_probability_expansion}), shown as solid lines. The sum of the LNC and LNV probability ,i.e.\ $P_{\mu \Sigma}^{LNC} (L) + P_{\mu \Sigma}^{LNV} (L)$ is identical zero. The legend labels lines of increasing amplitude from left to right.}\label{fig:mixing_probability_NLO_muSigma}
    \end{subfigure}
    \end{minipage}
    }
\caption{NLO contributions to the heavy neutrino-antineutrino oscillation probabilities.}
\end{figure}

Using the results from \cref{sec:lowscaleseesaw} the oscillation probability \cref{eq:probability_LNX} can be expanded in the small parameters $\vec{\theta}'$ appearing in the lepton mixing matrix and $\lambda_m$ which has been introduced in \cref{eq:def_masssplitting_parameter} to account for the mass splitting of the heavy neutrinos. Also, as we discussed, the mass splitting parameter $\lambda_m$ can be neglected at leading order as it is much smaller than $\mathcal{O}(\vec{\theta}')$, see \cref{eq:approx_lambdam}. The following definitions are used
\begin{align*}
I_\beta &:= \mathrm{Im}(\theta_\beta^* \theta'_\beta \exp(-2 i \Phi)) \:,\\
\phi_{ij} &:=- \frac{2 \pi}{L_{ij}^{osc}} = -\frac{m_i^2 - m_j^2}{2 |\mathbf{p}_0|}\:,\\
\Phi &:= \frac{1}{2}\mathrm{Arg}\left(\vec{\theta}' \cdot \vec{\theta}^*\right)\:.
\end{align*} 
To derive the approximate oscillation formulae, the normalisation constant is computed using the condition \cref{eq:normalisation_condition}, yielding
\begin{equation}
N_g^2 = \frac{1}{\sum_\beta |\theta_\alpha|^2 |\theta_\beta|^2 + \dots~},
\end{equation}
where the ellipses contain orders higher then $\mathcal{O}(\vec{\theta}')$. With this the probability \cref{eq:probability_LNX} can be expanded in $\vec{\theta}'$. Note that we keep the exponential $\exp \left( i \Phi_{ij} L \right)$ exact. Up to first order in $\vec{\theta}'$, the probability \cref{eq:probability_LNX} in the LNC case yields
\begin{equation}
\label{eq:LNC_probability_expansion}
\begin{split}
P^{LNC}_{\alpha \beta}(L) &= \frac{1}{2 \sum_\beta |\theta_\alpha|^2 |\theta_\beta|^2} \bigg( |\theta_\alpha|^2 |\theta_\beta|^2 (1 + \cos (\phi_{45} L)) \\
&\quad - 2 (I_\beta |\theta_\alpha|^2  - I_\alpha |\theta_\beta|^2 ) \sin (\phi_{45} L) \bigg)~.
\end{split}
\end{equation}
In the LNV case the expansion of \cref{eq:probability_LNX} up to first order in $\vec{\theta}'$ yields
\begin{equation}
\label{eq:LNV_probability_expansion}
\begin{split}
P^{LNV}_{\alpha \beta}(L) &= \frac{1}{2 \sum_\beta |\theta_\alpha|^2 |\theta_\beta|^2} \bigg( |\theta_\alpha|^2 |\theta_\beta|^2 (1 - \cos (\phi_{45} L)) \\
&\quad - 2 (I_\beta |\theta_\alpha|^2  + I_\alpha |\theta_\beta|^2 ) \sin (\phi_{45} L) \bigg)~,
\end{split}
\end{equation}
where in both cases the LO terms are written in the first line and the NLO terms in the second line. Note that if the initial $W$ boson is replaced by a $W^-$, the leptonic mixing matrix factors are complex conjugate to the ones in the process where the initial boson is a $W^+$. This results in a sign change of the NLO contributions. The leading order term in those expansions describe the oscillations from neutrinos into antineutrinos, whereas the first order term describes flavour oscillations. This can be seen by adding up the LNC and LNV probabilities, which means that the sign of the outgoing lepton is ignored. This should make the oscillations of neutrinos into antineutrinos vanish, and indeed the oscillatory part of the leading order terms cancel each other. We are left with
\begin{equation}
P^{LNC}_{\alpha \beta}(L) + P^{LNV}_{\alpha \beta}(L) = \frac{1}{ \sum_\beta |\theta_\alpha|^2 |\theta_\beta|^2}\bigg(  |\theta_\alpha|^2 |\theta_\beta|^2 - 2 I_\beta |\theta_\alpha|^2 \sin (\phi_{45} L) \bigg)~.
\end{equation}
Summing over the outgoing flavours makes the oscillatory part of the above equation vanish. This happens because $\sum_\beta \theta_\beta^* \theta'_\beta \exp(-2 i \Phi) \in \mathbb{R}$, and therefore $\sum_\beta I_\beta = 0$.
\\\\
With the ``lepton number''-like symmetry being broken, one also expects lepton number violation in the limit where the distance $L$ goes to zero. As mentioned above, the no-dispersion regime, which is the relevant one in this limit, results in the same formulae for the oscillation probabilities if the observability conditions are met. 
From \cref{eq:LNV_probability_expansion} it can therefore be seen that there is no lepton number violation at zero distance. This leads to the conclusion that this effect has to be introduced at a higher order and is therefore much smaller than the lepton number violation due to oscillations. That this effect is indeed present at higher orders can be confirmed by numerically diagonalizing the lepton mass matrix and using \cref{eq:probability_LNX} to compute the oscillation probability.
\\\\
Taking only the leading order into account, the only relevant model parameters are the Yukawa couplings $Y_\alpha$, or equivalently the mixing parameters $\theta_\alpha$, and the quadratic mass splitting $\delta m_{45}^2$ appearing in $\phi_{45}$. The leading order effects can therefore be described by the symmetry limit of the SPSS \cite{Antusch:2015mia,Antusch:2016ejd} plus the quadratic mass splitting as an additional parameter. This holds even more generally, for all realisations of the SPSS.
\\\\
If the mechanism of light neutrino masses is given by the minimal linear seesaw described in \cref{eq:lagrangian}, one can reparameterize the model in terms of active neutrino parameters according to \cite{Gavela:2009cd}. Assuming an inverse ordering of the light neutrino masses $m_{\nu_i}$ yields
\begin{equation}
Y_\alpha = \frac{y}{\sqrt{2}} \sqrt{1 + \rho}~U_{\alpha 2}^* + \sqrt{1-\rho}~U_{\alpha 1}^*
\end{equation}
\begin{equation}
 Y'_\alpha = \frac{ y'}{\sqrt{2}} \sqrt{1 + \rho}~U_{\alpha 2}^* - \sqrt{1-\rho}~U_{\alpha 1}^*
\end{equation}
where $U$ denotes the unitary PMNS matrix,
\begin{equation}
    \rho = \frac{\sqrt{1+r}-1}{\sqrt{1+r}+1}
\end{equation}
and
\begin{equation}
    r = \frac{|m_{\nu_1}^2 - m_{\nu_2}^2|}{|m_{\nu_1}^2 - m_{\nu_3}^2|}~.
\end{equation}
Note that we have absorbed the parameter $\epsilon$ appearing in \cite{Gavela:2009cd} into the definitions of $Y'$ and $y'$. An interesting observation is that the heavy neutrino mass splitting is given by the light neutrino mass splitting \cite{Antusch:2017ebe}
\begin{equation}
 m_{5} - m_{4} = m_{\nu_2} - m_{\nu_1}~,
\end{equation}
and therefore the squared mass splitting is given by
\begin{equation}
\label{eq:squared_mass_splitting_ligntnu}
    \delta m_{45}^2 = - 2 \tilde{m}_0 (m_{\nu_2} - m_{\nu_1})~.
\end{equation}
Taking the values of the active neutrino mixing angles and mass squared differences from \cite{Esteban:2018azc,Nufit:2019}, the only undetermined parameter is the Majorana phase $\alpha$. Analyzing the parameterization of the Yukawa couplings one finds that the products $Y_\alpha Y_\alpha^*,~{Y'}_\alpha {Y'}_\alpha^*,~{Y'}_\alpha Y_\alpha^*$ are only independent of the Majorana phase when summed over the flavour index. The oscillation probabilities \cref{eq:LNC_probability_expansion,eq:LNV_probability_expansion} are visualized in \subref{fig:mixing_probability_LO_mumu}, \subref{fig:mixing_probability_LO_muSigma}, \subref{fig:mixing_probability_NLO_mumu} and \subref{fig:mixing_probability_NLO_muSigma}. The values of the oscillation probabilities makes it clear that any higher order effects on the oscillation patters will be extremely challenging to observe under realistic conditions for the chosen example parameters from \cref{table:parameter_simulation}.

\section{Summary and Conclusions}
In this paper we have applied the framework of quantum field theory (QFT) with external wave packets  (cf.~\cite{Beuthe:2001rc}) to derive the probabilities for the oscillations between long-lived heavy neutrino and antineutrino interaction eigenstates (cf. \cref{fn:definition_heavy_neutrino_interaction_eigentate}), where we define a neutrino (antineutrino) as the neutral lepton that is produced together with a charged antilepton (lepton) and a $W$ boson. These heavy neutrino-antineutrino oscillations can lead to an oscillating rate of lepton number conserving (LNC) and violating (LNV) events at colliders, as a function of the distance between the (anti)neutrino production and displaced decay vertices. 
\\\\
Our most general formula for the oscillation probability is \cref{eq:probability_LNX_general} together with the additional terms discussed in \cref{subsec:observability_conditions}. The latter can be neglected given the adequate kinematic and experimental conditions, and are referred to as observability conditions. The oscillation probability densities can be further simplified to \cref{eq:probability_LNX} by neglecting the spin correlation between the production and detection vertex. Including the decay probabilities of the heavy neutrinos, formulae for the expected number of LNC/LNV events with a certain displacement between primary and secondary vertex have been given in \cref{eq:number_of_events}.
\\\\
The simplified formulae for the oscillation probabilities have been applied to low scale seesaw models where the smallness of the light neutrino masses is protected by a slightly broken ``lepton number''-like symmetry, i.e.\ to the SPSS (cf.\ \cite{Antusch:2015mia,Antusch:2016ejd}). As a particular example we have focused on the ``minimal low scale linear seesaw'' model with only two nearly mass-degenerate heavy neutrinos, that has also been discussed as an example in \cite{Antusch:2017ebe}. 
Within this class of models, an expansion of the probabilities in the small ``lepton number''-like symmetry breaking parameters $\vec{\theta}'$ has been performed, yielding the LO and NLO contributions (cf.\ \cref{eq:LNC_probability_expansion,eq:LNV_probability_expansion}).
\\\\
For the example parameter point used in Ref.~\cite{Antusch:2017ebe}, we have discussed the observability conditions (cf.\ \cref{subsec:observability_conditions}) and found that they are all satisfied. However, if the momentum $\mathbf{p}_0$ is given by a distribution, which is the case if the mean momenta of the external particles follow a distribution as discussed in \cref{subsec:observability_conditions}, the oscillation pattern can be washed out as has already been pointed out in \cite{Antusch:2017ebe}. The proposed solution to this is to reconstruct the four momentum $p_0$ (from the measurements) and to consider the oscillations as a function of the heavy neutrino reconstructed proper time. In \cite{Antusch:2017ebe} it has been demonstrated, using estimated uncertainties for the HL-LHCb and the above-mentioned example parameter point, that the proposed solution is indeed feasible.  
\\\\
Comparing our simplified LO results with the existing literature, we found that we agree with the results from \cite{Cvetic:2015ura} (when we set their parameters $\theta_{21}^{LNV} = - \pi$ and $\theta_{21}^{LNC} = 0$ to match the considered low scale seesaw scenario). Our LO formulae also agree with the ones derived using the formalism for meson oscillations and plane wave arguments and used e.g.\ in \cite{Anamiati:2016uxp,Antusch:2017ebe}. Our results in the most general form, i.e.\ \cref{eq:probability_LNX_general} together with the additional terms discussed in \cref{subsec:observability_conditions}, allow to discuss effects beyond LO and to check the observability conditions (or include them explicitly in the calculations).
\\\\
Our NLO results showed that beyond the LO heavy neutrino-antineutrino oscillations, the probablities are also modulated by ``flavour oscillations'', as discussed in \cref{sec:approximations_oscillation_formula}. On the other hand, for the case of the ``minimal low scale linear seesaw'' model (with parameters around the considered example point), it has turned out that the NLO effects are very small, with a suppression which makes them undetectable at the currently considered future collider experiments. While this does not necessarily have to be the case for other choices of parameters, it indicates that there is a parameter region of interest for the LHC (and future colliders) where the LO formulae are sufficient. In this region the only model parameters relevant for heavy neutrino-antineutrino oscillations, in terms of the SPSS parameters, are the three flavour-dependent active-sterile mixing angles $\theta_\alpha$ and the mass squared difference $\delta m_{45}^2$ between the two heavy neutrinos.     
\\\\
In summary, our results show that the phenomenon of heavy neutrino-antineutrino oscillations can indeed occur in low scale seesaw scenarios and that the previously used leading order formulae, derived with a plane wave approach, provide a good approximation for (at least) the considered example parameter point. Our results help to put existing studies based on LO formulae on a more solid theoretical ground (by providing the observability conditions which have to be checked in the QFT framework) and can be used in future studies to explore the phenomenon in other parameter regions and for different types of low scale seesaw models.

\section*{Acknowledgements}
We thank Joachim Kopp for helpful discussions on the QFT formalism for neutrino oscillations. This work has been supported by the Swiss National Science Foundation (project number 200020/175502).

\appendix

\section{Formulas for the Observability Conditions}
\label{apx:formulas_observability_conditions}
In the following, the formulas to compute the observability conditions of \cref{subsec:observability_conditions} are given. For more details we refer to \cite{Beuthe:2001rc}.
\\\\
The particles at production are labeled $P_i$ and their wave packets are assumed to have a Gaussian form of width $\sigma_{x P_i}$ in configuration space. The wave packet peaks, in momentum space, at momentum $\mathbf{P}_i$ and since they are assumed to be on-shell, their peak energy is given by $E_{P_i} = \sqrt{m_{p_i}^2 + |\mathbf{P_i}|^2}$. The peak velocity is then defined as $\mathbf{v}_{P_i} = \mathbf{P}_i/E_{P_i}$. For the particles at detection the letter $P$ is simply replaced by $D$. The velocity $\mathbf{v}_0$ is defined using energy-momentum conservation at the production and/or detection vertex, which yields 
$$\mathbf{v}_0 := \mathbf{p}_0/ E_0~.$$
Labelling the incoming particles at production $P_{i, in}$ and the outgoing ones $P_{i, out}$ yields
\begin{equation}
\label{eq:apx_E0}
    E_0 := \sum_{P_{i, in}} E_{P_{i, in}} - \sum_{P_{i, out}} E_{P_{i, out}}
\end{equation}
and 
\begin{equation}
\label{eq:apx_p0}
  \mathbf{p}_0 := \sum_{P_{i, in}} \mathbf{P}_{i, in} - \sum_{P_{i, out}} \mathbf{P}_{i, out}~.  
\end{equation}
The relevant parameters are defined as follows.
\begin{equation}
\label{eq:apx_sigmaxPD}
    \frac{1}{\sigma_{x P}^2} = \sum_{P_i} \frac{1}{\sigma_{x P_i}^2} \qquad \frac{1}{\sigma_{x D}^2} = \sum_{D_i} \frac{1}{\sigma_{x D_i}^2}
\end{equation}

\begin{equation}
\label{eq:apx_sigmapPD}
    \sigma_{p P} = \frac{1}{2 \sigma_{x P}} \qquad \sigma_{p D} = \frac{1}{2 \sigma_{x D}}
\end{equation}

\begin{equation}
\label{eq:apx_vPD}
    \mathbf{v}_P = \sigma_{x P}^2 \sum_{P_i} \frac{\mathbf{v}_{P_i}}{\sigma_{x P_i}}\qquad  \mathbf{v}_D = \sigma_{x D}^2 \sum_{D_i} \frac{\mathbf{v}_{D_i}}{\sigma_{x D_i}}
\end{equation}

\begin{equation}
\label{eq:apx_SigmaPD}
    \Sigma_P = \sigma_{x P}^2 \sum_{P_i} \frac{|\mathbf{v}_{P_i}|^2}{\sigma_{x P_i}} \qquad \Sigma_D = \sigma_{x D}^2 \sum_{D_i} \frac{|\mathbf{v}_{D_i}|^2}{\sigma_{x D_i}}
\end{equation}

\begin{equation}
\label{eq:apx_sigmaePD}
    \sigma_{e P}^2 = \sigma_{p P}^2 \left( \Sigma_P - |\mathbf{v}_P|^2 \right) \qquad \sigma_{e D}^2 = \sigma_{p D}^2 \left( \Sigma_D - |\mathbf{v}_D|^2 \right)
\end{equation}

\begin{equation}
\label{eq:apx_sigmap}
    \frac{1}{\sigma_p^2} = \frac{1}{\sigma_{p P}^2} + \frac{1}{\sigma_{p D}^2}
\end{equation}

\noindent The following symbols are defined in the longitudinal dispersion regime, on which we focused in this paper. We denote a velocity ($\mathbf{v}$) projected onto the direction in which the oscillation distance is measured ($\hat{\mathbf{L}}$) as $\nu$.\footnote{If one does only measure the scalar distance of oscillation, one might replace the direction $\mathbf{L}/|\mathbf{L}|$ with the direction $\mathbf{p}_0/|\mathbf{p}_0|$, which can be interpreted as the main direction in which the heavy neutrinos travel.}

\begin{equation}
    \frac{1}{\sigma_{p eff}^2}  = \frac{1}{\sigma_{p P}^2} + \frac{1}{\sigma_{p D}^2} + \frac{(\nu_0 - \nu_P)^2}{\sigma_{e P}^2} + \frac{(\nu_0 - \nu_D)^2}{\sigma_{e D}^2}
\end{equation}

\begin{equation}
    \sigma_{x eff} = \frac{1}{2 \sigma_{p eff}}
\end{equation}

\begin{equation}
    \rho = \sigma_{p eff}^2 \left( \frac{1}{\sigma_{p P}^2} + \frac{1}{\sigma_{p D}^2} - \frac{\nu_P (\nu_0 - \nu_P)}{\sigma_{e P}^2} - \frac{\nu_D (\nu_0 - \nu_D)}{\sigma_{e D}^2}\right)
\end{equation}

\begin{equation}
    \frac{1}{\sigma_m^2} = \sigma_{p eff}^2 \left( \frac{1}{\sigma_p^2} \left(\frac{1}{\sigma_{e P}^2} + \frac{1}{\sigma_{e D}^2}\right) + \frac{(\nu_P - \nu_D)^2}{\sigma_{e P}^2 \sigma_{e D}^2} \right)
\end{equation}

\newpage
\bibliographystyle{unsrt}

\end{document}